\documentclass[aps,showpacs,notitlepage,reprint,onecolumn,12pt,preprintnumbers,amsmath,amssymb,nofootinbib]{revtex4-1}
\usepackage{graphicx}
\usepackage{dcolumn}
\usepackage{bm}
\usepackage{setspace}
\usepackage{amsmath}
\usepackage{enumerate}

\def\simge{\mathrel{
     \rlap{\raise 0.511ex \hbox{$>$}}{\lower 0.511ex \hbox{$\sim$}}}}
\def\simle{\mathrel{
     \rlap{\raise 0.511ex \hbox{$<$}}{\lower 0.511ex \hbox{$\sim$}}}}
\def\be{\begin{equation}}
\def\ee{\end{equation}}
\def\bea{\begin{eqnarray}}
\def\eea{\end{eqnarray}}

\newcommand{\calh}{{\cal H}}

\newcommand{\calo}{{\cal O}}

\newcommand{\Tr}{{\rm Tr}\,}
\newcommand{\ReTr}{{\rm ReTr}\,}
\renewcommand{\Re}{{\rm Re}\,}

\newcommand{\ax}{a_x}
\newcommand{\axd}{a_x^\dagger}
\newcommand{\bx}{b_x}
\newcommand{\bxd}{b_x^\dagger}
\newcommand{\ay}{a_y}

\newcommand{\by}{b_y}

\begin{document}
\singlespacing
\title{Monte-Carlo simulation of the tight-binding model of graphene 
with partially screened Coulomb interactions}
\author{Dominik Smith,$^{a}$ Lorenz von Smekal$^{a,e}$}
\affiliation{
$^a$Theoriezentrum, Institut f\"ur Kernphysik, TU Darmstadt, 64289 Darmstadt, Germany\\ 
$^e$Institut f\"ur Theoretische Physik, Justus-Liebig-Universit\"at,
   35392 Giessen, Germany 
}
\date{\today}
\begin{abstract}
We report on Hybrid-Monte-Carlo simulations of the tight-binding
model with long-range Coulomb interactions for the electronic
properties of graphene. 
We investigate the spontaneous breaking of sublattice symmetry
corresponding to a transition from the semimetal to an
antiferromagnetic insulating phase.
Our short-range interactions thereby include the partial screening due
to electrons in higher energy states from ab initio
calculations based on the constrained
random phase approximation [T.~O.~Wehling {\it et al.},
    Phys.~Rev.~Lett.~{\bf 106}, 236805 (2011)].
In contrast to a similar previous Monte-Carlo study [M.~V.~Ulybyshev
  {\it et al.}, Phys.~Rev.~Lett.~{\bf 111}, 056801 (2013)] we also
include a phenomenological model which describes the transition
to the unscreened bare Coulomb interactions of graphene at half filling 
in the long-wavelength limit. Our results show, however, 
that the critical coupling for the antiferromagnetic Mott transition
is largely insensitive to the strength of these long-range Coulomb
tails. They hence confirm the prediction that suspended graphene
remains in the semimetal phase when a realistic static screening
of the Coulomb interactions is included.
\end{abstract}
\maketitle

\section{Introduction}

In recent years much interest has arisen in the study of
graphene, an allotrope of carbon which consists of a single
layer of atoms, arranged on a hexagonal (``honeycomb'') lattice. 
It has become increasingly clear that such a system, despite
(or rather due to) its very simple structure, possesses a large variety
of unusual properties. These range from extreme mechanical
strength and lightness, over unique electronic properties
to a number of anomalous quantum effects, which make 
graphene a very attractive candidate for a wide range of technological
applications (for extensive reviews of the properties of graphene,
see Refs.~\cite{Geim:2007jh,CastroNeto:2009zz,Kotov:2012ff,Beenakker:2008df}). 
Moreover, from a theoretical perspective
it has become clear that graphene can serve as a model system
for a large number of concepts from high-energy physics, ranging
from topological phase transitions, chiral symmetry breaking
and super-symmetry to quantum gravity (for more on the
connections between graphene and high-energy physics see 
e.g.~Refs.~\cite{Novoselov:2005df,Zhang:2005ht,Semenoff:2011sd,Gusynin:2007ix,Park:2010zs,Abreu:2010yv,Cortijo:2007fd,Iorio:2013ifa}
and references therein).

This has motivated
the application of well-established field theory methods from particle 
physics as effective descriptions of the low-energy electronic excitations
in graphene. Since the relevant coupling constant for the Coulomb
interactions can thereby be as large as
$\alpha_\mathrm{eff}=e^2/(\hbar v_F)$ in suspended graphene, 
where $v_F\approx c/300$ is the Fermi velocity, one needs to describe 
a strongly-coupled fermionic system  with an effective coupling
$\alpha_\mathrm{eff}\approx 300/137\approx 2.2$. This suggests
the application of non-perturbative methods.
In particular, graphene at half filling can be efficiently simulated via 
Hybrid-Monte-Carlo, a widely used algorithm in
lattice gauge theory, both in the low-energy
(long-wavelength) limit  
\cite{Drut:2008rg,Drut:2009gg,Drut:2009rf,Drut:2010ff,Drut:2011fd,Hands:2008dg,Armour:2009vj,Armour:2011ff,Buividovich:2012kk,Giedt:2011df}
where graphene is well-described by a variant of
Quantum Electrodynamics in $2+1$ dimensions,
and as a full theory which is valid on length-scales
down to the interatomic distance $a\approx 1.42 $ \AA\ 
\cite{Brower:2012zd,Brower:2012ze,Buividovich:2012nx,Ulybyshev:2013swa,Smith:2013pxa}.

An open question which is of immediate consequence to 
technological applications is whether graphene, which is
known to be an electric conductor when affixed to a
number of different substrates, can develop 
a band gap under proper circumstances. 
This could correspond to a spontaneous breaking of the symmetry under
exchange of the two triangular sublattices 
with strong analogies to chiral symmetry breaking in relativistic
field theory.
A substrate generates dielectric screening which
lowers the effective fine-structure constant $\alpha_\mathrm{eff}$
of the system. The expectation is that if
screening is reduced, when $\alpha_\mathrm{eff}$ becomes larger than
some critical coupling $\alpha_c$, a phase transition
to a gapped phase occurs. In order to be physically realizable,
$\alpha_c$ should be smaller than
$\alpha_{\textrm{eff},0}\approx 300/137 \approx 2.2$ 
which is the upper bound in suspended graphene,
where screening is minimal.

Experiments have provided evidence 
that graphene in vacuum is in fact a conductor
\cite{Elias:2011gz,Mayorov:2012sf},  
while analytical calculations 
\cite{Gamayun:2010gd,Leal:2004df,Araki:2010ss,Araki:2011hf,Araki:2012hj} 
and simulations 
\cite{Drut:2008rg,Drut:2009gg,Drut:2009rf,Hands:2008dg,Armour:2009vj,Buividovich:2012nx}, which
assumed that the electromagnetic interactions of 
$\pi$-band electrons (the relevant degrees of freedom
for the electronic properties) 
are essentially unmodified Coulomb interactions, supported the scenario
of a gapped phase for $\alpha_\mathrm{eff} > \alpha_c\approx 1$, well within
the accessible region. The origin of this disparity must thus be
investigated. 

Recently, it was suggested that additional screening
(independent of the reduction of $\alpha_\mathrm{eff}$ by the
substrate) of the
two-body Coulomb interactions, by electrons in the
lower $\sigma$-bands and other higher energy states 
of the carbon sheet itself \cite{Wehling:2011df}, provides a mechanism
which moves $\alpha_c$ to larger values, outside of the
physical region.\footnote{Another mechanism which
has been proposed is a reshaping of the 
Dirac cone due to renormalization of the Fermi velocity 
\cite{Elias:2011gz,Popovici:2013fwp,Popovici:2013wfa}. 
The magnitude of this effect is likely much smaller than
that of screening and, in any case, the inclusion of
this effect is automatic in lattice simulations.} In
Ref.~\cite{Ulybyshev:2013swa} 
Hybrid-Monte-Carlo simulations of the tight-binding
model with an instantaneous two-body potential
generated by a Hubbard field were carried
out (based on the framework developed in
Refs.~\cite{Brower:2012zd,Brower:2012ze})  
which addressed the issue of $\sigma$-band screening. 
For these
simulations a screened Coulomb potential was chosen which
used the results of a calculation within the 
constrained random phase approximation (cRPA)
\cite{Wehling:2011df}
for on-site repulsion, the nearest-neighbor, next-to-nearest-neighbor
and third-nearest-neighbor interactions at short distances.
At longer distances it was assumed that the potential falls
off as $\sim 1/(\epsilon_\sigma \, r)$, where the constant
$\epsilon_\sigma \approx 1.41 $ was adjusted to match the third-nearest-neighbor
term. It was shown that for this particular choice of potential, the
critical coupling for the antiferromagnetic Mott transition 
is moved to $\alpha_c\approx 3.14$, which is
outside of the physically accessible region and thus agrees
with the experimental observation. 

In this work, we conduct simulations similar to those of
Ref.~\cite{Ulybyshev:2013swa}, however with a more realistic 
description of the partial screening of the Coulomb interactions 
at larger distances: Instead of assuming the constant reduction in the 
strength of the long-range Coulomb tails, 
by $\epsilon_\sigma $ which quite naturally necessitates an increased effective
coupling to compensate that, we use the phenomenological model also
given in Ref.~\cite{Wehling:2011df} to construct a 
partially screened Coulomb interaction with a momentum dependent
$\epsilon_\sigma(\vec k)$ which smoothly turns into
the unscreened Coulomb potential corresponding to $\epsilon_\sigma\to
1$ in the long-wavelength limit. This reflects the fact that
the high energy states in graphene do not screen the long-range
Coulomb tails in the interactions of the $\pi$-band electrons
as demonstrated explicitly in \cite{Wehling:2011df}. Because the
density of states in the $\pi$-bands furthermore 
vanishes at the Dirac points in the band structure of graphene, so
does the static Lindhard susceptibility in the long-wavelength
limit. There is thus no screening of the long-range Coulomb tails from
the $\pi$-bands in graphene at half filling in the semimetal phase
either. The presence of unscreened long-range Coulomb interactions is
one of the distinctive features of the interactions in graphene.

The effects of on-site repulsion versus nearest and few
next-to-nearest-neighbor interactions have been studied in
Hubbard models at length. Depending on their relative strengths a 
variety of competing phases has been predicted such as spin and
charge-density-wave phases
\cite{Sorella:1992,Herbut:2006,Honerkamp:2008}, spin liquids
\cite{Meng:2010,Tran:2011} or  topological insulators
\cite{Raghu:2008}. In order to address the long-standing 
question whether the long-range Coulomb interactions in graphene favor
any of these insulating phases we therefore decided to 
rather leave those Coulomb tails unmodified. The fact that our results
basically agree with those of Ref.~\cite{Ulybyshev:2013swa} indicates
that they have little effect on the antiferromagnetic
spin-density-wave formation investigated here.   

This paper is structured as follows: In the next section we present
a detailed review of the theoretical framework for the 
Hybrid-Monte-Carlo simulation of the hexagonal graphene lattice
which is employed in this work, including a discussion of the 
partially screened Coulomb potential which is used. We then present
the results of the investigation of the semimetal-insulator phase transition,
including a more detailed comparison with the results of
Ref.~\cite{Ulybyshev:2013swa} which we have reproduced for cross-checking
purposes. In the last section, we discuss our results and provide
an outlook on future projects. 

A few comments on the units and conventions used throughout
this paper: We use the natural system of units of high-energy physics,
i.e. $c\equiv 1$ and $\hbar\equiv 1$. Furthermore, we use the Gaussian 
system of electromagnetic
units, in which the Coulomb potential is simply $1/r$. 
These conventions
imply the relation $\alpha=e^2 \approx 1/137$ between
the elementary charge and the (dimensionless) fine-structure constant.
We then have the freedom to express all physical quantities
using either a basic unit of energy or of length. We generally
use electron-volt ($\mathrm{eV}$) as the fundamental unit 
(the potential then has a dimension of energy). However,
in some cases (where experimental results are concerned) 
a value in meters will be also given for lengths.
The two are related via 
$10^{-7}\mathrm{m}\approx 0.506\, \mathrm{eV}^{-1}$.
When conducting Fourier analysis we write 
normalization factors
of $1/(2\pi)$ only in front of momentum-space
integrals, as is customary in high-energy physics.

\section{The Setup}
In this section the central components of the lattice
simulation of the interacting tight-binding theory of
graphene are derived. We aim to give a rather comprehensive
review, which should hopefully be useful for readers from
various backgrounds. We therefore provide
many technical details, and where we refer to existing literature we
fill in various steps that where omitted in the original articles. 

The section is structured
as follows: We begin with a general discussion of the graphene lattice
and of the way we treat boundary conditions, followed by 
the derivation of the path-integral
formulation of the grand-canonical partition function $Z$
of the interacting tight-binding
theory. 
This forms the basis of everything that follows.
The derivation of $Z$ was first worked out in
Refs.~\cite{Brower:2012zd,Brower:2012ze}. We review the
essential steps and provide an explicit representation
of the fermion matrix.
Based on this formulation, in the following paragraph a detailed
discussion of the Hybrid-Monte-Carlo algorithm is then presented,
which we use to generate lattice configurations. This includes
a step-by-step derivation of the molecular-dynamics force terms,
which have not been given explicitly in the literature before.
Next, we present the derivation of an expression for the
order parameter for sublattice symmetry breaking in
terms of elements of the inverse fermion matrix, which 
is used for measurements. We then discuss the 
second order fermion operator first derived in
Ref.~\cite{Ulybyshev:2013swa}, specifically showing
which aspects of the preceding derivations are changed.
And finally we discuss in detail
how we obtain an expression for the partially screened
Coulomb potential using the cRPA results
and the phenomenological dielectric
screening function presented in Ref.~\cite{Wehling:2011df}.

\subsection{The graphene lattice}
\label{subsec:graphlat}
Consider a 
two-dimensional triangular lattice, spanned
by the basis vectors
\be
\vec{e_1}=(\sqrt{3}, 0)\,a~,\quad \vec{e_2}=(\frac{\sqrt{3}}{2}, \frac{3}{2})\,a~,
\ee
such that each lattice point can be reached by 
$\vec{r}=m \vec{e_1}+n \vec{e_2}$ for some integers $m,n$.
Here we have introduced 
$a=\,1.42\; \textrm{\AA}$ ($\approx 0.71\cdot 10^{-3}\,\textrm{eV}^{-1}$). 
The hexagonal graphene lattice
can be constructed from these vectors by assigning a two-component
basis to each lattice point, such that one carbon atom sits exactly
on each point and another one is reached by a translation 
\be
\vec{\delta}= (0, 1)\,a~.
\ee
It is obvious from this construction that $a$ is the interatomic spacing.
The above is equivalent to the statement that graphene is composed of
two inequivalent triangular sublattices, which sit a translation
along $\vec{\delta}$ apart. We will refer to these as 
sublattices $A$ and $B$. 

By restricting $(m,n)$ to $m\in [0,L_m-1]$, $n\in [0,L_n-1]$ one obtains
a graphene sheet shaped like a parallelogram. In the following we always assume
that $L_m,L_n$ are both even. It is our goal to simulate rectangular
graphene sheets with periodic boundary conditions. We thus impose 
\be
(m+L_m,n)\equiv (m,n)~,\quad (m,n+L_n)\equiv (m-L_n/2,n)~.
\ee
The periodic boundary conditions, invariant under discrete hexagonal
lattice translations,  are a technical device to
reduce boundary effects. As such they are frequently used in lattice
simulations when one is (as we are here) interested in
bulk thermodynamics. Their main purpose is not so much to mimic
physical boundary conditions in experiments but to provide a reasonably rapid
approach towards the infinite volume limit with growing system
size. 

To construct a finer rectangular system of coordinates whose basis
vectors align with the axes of periodicity, and in which
points on both sublattices can be uniquely identified,
consider the new set of basis vectors given by
\be
\vec{e_1}'=\vec{e_1}/2 =  (\sqrt{3}/2, 0)\,a~,
\quad\vec{e_2}'= \vec{\delta}/2 = (1/2,0)\,a~.
\ee
The majority of points on the finer grid defined by these vectors are
empty, but every point on the hexagonal lattice can be 
written as $\vec{r}\,'=x \vec{e_1}'+y \vec{e_2}'$ with 
\be
x=2m+n~,\quad y = 3n+2 P_{AB}~,
\ee
where $P_{AB}=0$ on  sublattice $A$ and $P_{AB}=1$ on  sublattice
$B$. Periodicity is expressed in 
this system by restricting $x,y$ 
to $x\in [0,2L_m-1]$, $y\in [0,3L_n-1]$
and identifying
\be
(x+2L_m,y)\equiv (x,y)~,\quad (x,y+3L_n)\equiv (x,y)~.
\ee
The rectangular coordinate system $\vec{e_1}'$, $\vec{e_2}'$ will be
convenient for Fourier analysis.

\begin{figure}
\begin{center}
\resizebox{0.49\textwidth}{!}{%
\includegraphics{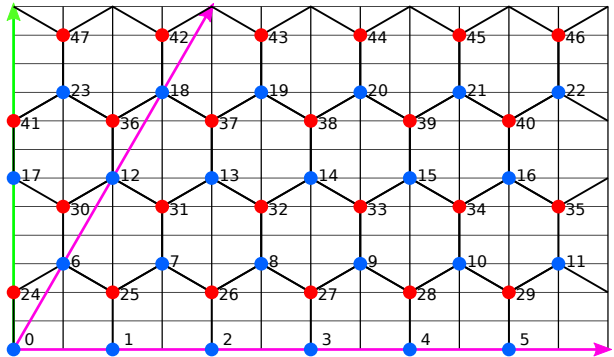}}
\end{center}
\caption{\label{fig:graphene_lattice}
Indexing scheme and coordinate systems for 
$L_m=6$, $L_n=4$, $N_t=1$. Blue dots are $A$-sites and
red dots are $B$-sites.}
\end{figure}

Fig. \ref{fig:graphene_lattice} shows an example in which 
$L_m=6$, $L_n=4$. The axes of both coordinate systems
are shown, including an indexing scheme (discussed below). 
The periodic system can be
constructed by repeating the figure.
The figure is drawn such that this
is well-defined: every lattice point on the boundary exists exactly once.

In the following paragraph an Euclidean (imaginary) time dimension
will be introduced. One can thus imagine additional
lattices associated with integer values of an additional
coordinate $\tau$. 
We store fields (functions defined over
the hexagonal lattice) as one-dimensional arrays. We thus
require a scheme to uniquely assign an index to each lattice
site. Consider a graphene lattice of spatial dimensions $L_m$ and $L_n$ 
with $N_t$ time slices. There are thus altogether  
$2\times L_m \times L_n \times N_t$
sites (the factor $2$ is due to the two sublattices).
We introduce lexicographically ordered site indices in such a way that
the entire $A$ sublattice is counted first, and the $B$ sublattice is counted
second. The index for a site ($m$,$n$,$\tau$) in one sublattice is thus
calculated  as
\be
\textrm{index}= m + n\, L_m + \tau\,(L_m\,L_n) + P_{AB}\,(L_m\,L_n\,N_t)~.
\ee
The example in Fig. \ref{fig:graphene_lattice} corresponds to $N_t=1$.

\subsection{Path integral formulation of the partition function}

We wish to express the grand-canonical partition 
function $Z=\Tr e^{-\beta H}$ of the tight-binding model
as a functional integral. This is essential for a Monte-Carlo
simulation, since in doing so one replaces operators by field
variables which can be stored in a computer. 
Here $\beta=1/k_B T$ (which we will express in units of
$\mathrm{eV}^{-1}$) refers to the temperature of the electron gas,
which does not include lattice vibrations and thus should not be 
confused with the temperature of the graphene sheet ($T$ can be
arbitrarily large here, whereas physical graphene would be
destroyed above some temperature).
We use the coherent state functional integral formalism (see
e.g.~Ref.~\cite{Negele:1988} for an introduction) to obtain $Z$. This formalism
derives from the fact that the Fock space of a fermionic many-body
system is spanned by a complete basis of fermionic coherent states.

The starting point is the Hamiltonian of the interacting tight-binding
model, 
\begin{align}
H &= H_{tb}+H_c+H_m \label{eq:Hamiltonian1}\\
&=-\kappa \sum_{\langle x,y \rangle,s}(a_{x,s}^\dagger a_{y,s}+a_{y,s}^\dagger a_{x,s}) 
+ \frac{1}{2} \sum_{x,y}\, q_x V_{xy} q_y +\sum_{x}m_s(a_{x,+1}^\dagger a_{x,+1}+a_{x,-1} a_{x,-1}^\dagger) ~,
\nonumber
\end{align}
where $a_{x,s}^\dagger, a_{x,s}$ are creation and annihilation operators
for electronic excitations in the $\pi$-bands with the 
usual fermionic anti-commutation relations,
\be
\{a_{i}^\dagger, a_{j}^\dagger\}=\{a_{i}, a_{j}\}=0~,\quad
\{a_{i}^\dagger, a_{j}\}=\delta_{ij}~.
\ee
The first sum in Eq.~(\ref{eq:Hamiltonian1})
runs over all pairs of nearest neighbors (including those pairs
where a periodic boundary is crossed) and defines the free
tight-binding Hamiltonian $H_{tb}$. The second and the
third sums run over all pairs of sites in the interaction Hamiltonian
$H_c$, and all sites in the ``mass term'' $H_m$, respectively.
The electron spin is labeled by $s=\pm 1$. The constant
$\kappa$ is the hopping parameter which is fixed by experiment to
$\kappa\approx 2.8\, \textrm{eV}$. The mass term is ``staggered'' 
with respect to the two independent sublattices which means that
\begin{equation}
   m_s  = \left\{ \begin{array}{lr}      
     +m\,, &  x \in A\, , \\
     -m\,, &  x \in B\, .  
      \end{array}
          \right.
\end{equation}       
This term is added to explicitly break sublattice symmetry. This is
required for technical reasons. In particular, it
removes zero modes from the fermion operator, such that
the fermion matrix can be inverted. Physical results are obtained
by extrapolating to $m \to 0$. Moreover note that the staggering in $H_m$ 
has the opposite sign for each spin component $s=\pm 1$,
and $m_s$ thus acts as an external field for spin-density-wave formation.

We have furthermore introduced
the charge operator
\be
q_x=a^\dagger_{x,1}a_{x,1}+a^\dagger_{x,-1}a_{x,-1}-1
=a^\dagger_{x,1}a_{x,1}- a_{x,-1} a^\dagger_{x,-1}~,
\ee
where the constant is added to make the system neutral at half filling.

$V_{xy}$ are the elements of a matrix describing 
instantaneous two-body interactions.
It need not be further specified at this point, other than that it
be positive-definite and have the dimension of energy (or inverse
length). We use a partially screened Coulomb potential 
in this work which is discussed in detail in
Sec.~\ref{subsec:partscreenedCoul} below.

To proceed we apply a few transformations to explicitly see how the
antiferromagnetic Mott transition at half filling can be simulated
without a fermion sign-problem\footnote{This refers to an indefinite
  sign or a complex measure introduced by the fermions which can make
  Monte-Carlo simulations impossible or at least extremely difficult.}
\cite{Sorella:1992}. The reason for this  
will become clear once we have obtained the explicit form of the fermion
operator. In particular, a mass term without the spin staggering, as
an external field for charge-density-wave formation, will be seen to
introduce such a sign problem.  
First, we introduce ``hole'' creation and annihilation operators
$b_{x}^\dagger, b_{x}$ for the electrons with $s=-1$ and from now on
use the notation
\be
a_x=a_{x,1}~,~a_x^\dagger=a_{x,1}^\dagger~,~
b_x^\dagger=a_{x,-1}~,~b_x=a_{x,-1}^\dagger~.
\label{eq:operators1}
\ee
The charge operator is then expressed as
\be
q_x = \axd \ax - \bxd \bx~,
\ee
and the Hamiltonian becomes
\begin{equation}
H = \sum_{\langle x,y \rangle}(-\kappa)(\axd \ay - \bxd \by + \textrm{h.c.}) 
+ \sum_{x,y}\, q_x V_{xy} q_y +  \sum_{x}m_s (\axd \ax + \bxd \bx)~.
\end{equation}
Next, we redefine the $b_{x}^\dagger, b_{x}$ by flipping the
sign on one of the sublattices, say $B$:
\be
\bx, \bxd \longrightarrow -\bx, -\bxd \quad \forall\, x \in B ~.
\ee
This is an allowed transformation since it preserves the anti-commutation
relations. It induces a sign-flip in the nearest-neighbor term:
\be
H_{tb} = \sum_{\langle x,y \rangle}(-\kappa)(\axd \ay - \bxd \by + \textrm{h.c.})
\longrightarrow \sum_{\langle x,y \rangle}(-\kappa)(\axd \ay + \bxd \by + \textrm{h.c.})~.
\ee

We now introduce the fermionic coherent states. These states are constructed
using anti-commuting Grassmann variables in the following way:

Consider a set of creation and annihilation operators 
$(a_\alpha,a^\dagger_\alpha )$, where the index $\alpha$ labels
all single-particle states of the system. Now associate with them
a set of Grassmann numbers $(\xi_\alpha,\xi^*_\alpha )$ such
that for each $a_\alpha$ there is a $\xi_\alpha$ and for each
$a_\alpha^\dagger$ there is a $\xi^*_\alpha$. The coherent states are
then
\be
|\xi \rangle= e^{-\sum_\alpha \xi_\alpha a_\alpha^\dagger  }|0\rangle~,\quad
\langle\xi|=\langle 0|\, e^{-\sum_\alpha a_\alpha \xi^*_\alpha }~.
\label{eq:coherent1}
\ee
The $|\xi \rangle$ are right-eigenstates of the annihilation operators,
while the $\langle\xi|$ are left-eigenstates of the creation operators.
The completeness relation on the Fock space is given in terms of these
states as 
\be
\int \left[ \prod_\alpha d\xi^{*}_\alpha \, d\xi_\alpha \right] \, e^{-\sum_\alpha \xi^{*}_\alpha \xi_\alpha}
|\xi\rangle \,\langle\xi| = {\bf 1}~,\label{eq:coherent_completeness1}
\ee
where the product and sum run over
all single-particle states. Moreover, 
the trace of a bosonic operator $A$ can be expressed as
\be                                                                             
\Tr A = \int                                                              
\left[ \prod_\alpha d\xi^{*}_\alpha \, d\xi_\alpha \right] \,
e^{-\sum_\alpha \xi^{*}_\alpha \xi_\alpha }
\langle -\xi|\,A\,|\xi \rangle~. \label{eq:coherent_trace}
\ee
We can now construct coherent states for the Fock space of fermionic
quasiparticles and holes, generated by the operators
$a_x,a^\dagger_x,b_x,b^\dagger_x$ as 
\be
\langle\psi,\eta|=\langle 0| e^{-\sum_x (\ax \psi_{x}^{*}+\bx \eta_{x}^{*})}~,\quad
|\psi,\eta\rangle= e^{-\sum_x (\psi_{x}\axd + \eta_{x}\bxd)}|0\rangle~.
\ee
Note that they are exactly of the form given in Eq.~(\ref{eq:coherent1}), if one understands
the index $\alpha$ to label spin and position states. Introducing two different symbols
$\psi,\eta$ is entirely a matter of notational convenience. Using these states, we can
now express the grand-canonical partition function as
\be                                                                                  
Z=\Tr e^{-\beta H} = \int                                                              
\left[ \prod_x d\psi^{*}_x \, d\psi_x \,d\eta^{*}_x \, d\eta_x \right] \,
e^{-\sum_x(\psi^{*}_x \psi_x +\eta^{*}_x \eta_x)}
\langle -\psi,-\eta|\,e^{-\beta H}\,|\psi,\eta\rangle~.\label{eq:tracecoher1}
\ee
In order to deal with the product
$\langle -\psi,-\eta|\,e^{-\beta H}\,|\psi,\eta\rangle$
which appears in the integral consider the
following: If $F(a^{\dagger}_\alpha,a_\alpha )$
is a \emph{normal ordered} function of creation and annihilation
operators, then its matrix element between coherent states is given
by
\be
\langle \xi |F(a^{\dagger}_\alpha,a_\alpha )|\xi' \rangle = 
F(\xi^{*}_\alpha,\xi'_\alpha )\, 
~e^{\sum_\alpha \xi_\alpha^{*} \xi_\alpha'}~.
\label{eq:matrixelement1}
\ee
Using our notation which distinguishes position and spin states
this becomes
\be
\langle\psi,\eta|F(a^{\dagger}_x,a_x,b^{\dagger}_x,b_x)|\psi',\eta'\rangle = 
F(\psi^{*}_x,\psi'_x,\eta^{*}_x,\eta'_x)\, 
~e^{\sum_x \psi_x^{*} \psi_x'+\eta_x^{*} \eta_x'}~.\label{eq:coher_theorem1}
\ee
If $e^{-\beta H}$ was a normal ordered function we would be done, for then
we could use Eq.~(\ref{eq:coher_theorem1}) and obtain an expression for $Z$
in which no operators appear. This is however not generally true, even if
$H$ is normal ordered. We proceed by splitting the exponential into
$N_t$ separate factors, 
\be
e^{-\beta H}=e^{-\delta H}\,e^{-\delta H}\ldots e^{-\delta H} \quad
(\delta=\beta/N_t)\, , 
\ee
and inserting a complete set of coherent states via the unity in
Eq.~(\ref{eq:coherent_completeness1}) 
between subsequent ones. Using an index $t$ to label the coherent states
associated with each time slice,
\be
\langle\psi_t,\eta_t|=\langle 0| e^{-\sum_x (\ax \psi_{x,t}^{*}+\bx \eta_{x,t}^{*})}~,\quad
|\psi_t,\eta_t\rangle= e^{-\sum_x (\psi_{x,t}\axd + \eta_{x,t}\bxd)}|0\rangle~,
\ee
we obtain the expression
\begin{align}
\Tr e^{-\beta H}=\int \prod\limits_{t=0}^{N_t-1} \left[\prod\limits_x
        d\psi^{*}_{x,t} \, d\psi_{x,t} \, d\eta^{*}_{x,t} \, d\eta_{x,t} \right]
        &~e^{-\sum_x( \psi^{*}_{x,t+1} \psi_{x,t+1}+\eta^{*}_{x,t+1} \eta_{x,t+1} )}\notag\\
        &\times\langle \psi_{t+1},\eta_{t+1}|e^{-\delta H}|\psi_{t},\eta_{t}\rangle~. 
\label{eq:partfunc}
\end{align}
From Eq.~(\ref{eq:partfunc}) it is clear that $t$ can be understood as
labeling the Euclidean time direction. Anti-periodic boundary conditions
\be
\psi_{x,N_t}=-\psi_{x,0}~,~\eta_{x,N_t}=-\eta_{x,0}~,
\ee
have been introduced here to make this compact notation possible.
They are simply the result of the minus sign in
Eq.~(\ref{eq:coherent_trace}), or inside the $\langle -\psi,-\eta|$
states in Eq.~(\ref{eq:tracecoher1}),  
and reflect the fermionic statistics of the electronic quasiparticles. 

The matrix elements in Eq.~(\ref{eq:partfunc}) are then
treated in the following way: Assuming that
the Hamilton operator is expressed in normal ordered form, the leading terms in
an expansion of $e^{-\delta H}$ which are not normal ordered are
$\calo(\delta^2)$. 
Discarding these terms, i.e. treating $e^{-\delta H}$ as if it was
normal ordered, therefore implies a discretization error $\calo(\delta)$ 
which vanishes with $N_t \to \infty$. Hence
$\delta$ can be visualized as the lattice spacing of the 
discretization in the Euclidean time direction.

We require a normal ordered form of $H$ to proceed. In this, the only
term requiring special attention is the diagonal term of $H_c$, since
this is the only term which generates something other than a trivial
change of sign when brought into normal order. Applying normal
ordering to this term yields:
\begin{align}
 q_x V_{xx} q_x & = - 2V_{xx} \, \axd \ax \, \bxd \bx +
 V_{xx} (\axd \ax + \bxd \bx)  =   \textbf{:} q_x V_{xx} q_x \textbf{:} +
{V_{xx}} (\axd \ax + \bxd \bx)~.
\end{align}
With $n_{x,\uparrow} = a_x^\dagger a_x$ and $n_{x,\downarrow} = 1 - b_x^\dagger
b_x$  one thus identifies the usual Hubbard model on-site
repulsion $U = V_{xx}$. 

Using Eq.~(\ref{eq:coher_theorem1}) we now
evaluate the matrix elements in Eq.~(\ref{eq:partfunc}) and obtain
\begin{align}
\Tr e^{-\beta H} &= \int \prod_{t=0}^{N_t-1} \left[ \prod_x d\psi^{*}_{x,t} \, d\psi_{x,t} \, d\eta^{*}_{x,t} \, d\eta_{x,t} \right]
\exp \Big\{ -\delta  \Big[\frac{1}{2}\sum_{x,y} Q_{x,t+1,t}V_{xy}Q_{y,t+1,t}\notag\\
&-\sum_{\langle x,y \rangle}\kappa
(\psi_{x,t+1}^{*} \psi_{y,t}+\psi_{y,t+1}^{*} \psi_{x,t}+\eta_{y,t+1}^{*} \eta_{x,t}+\eta_{x,t+1}^{*} \eta_{y,t}  )\notag\\
&+\sum_x
  m_s(\psi_{x,t+1}^{*}\psi_{x,t}+\eta_{x,t+1}^{*}\eta_{x,t})+\frac{1}{2}
  \sum_x V_{xx}(\psi_{x,t+1}^{*}\psi_{x,t}+\eta_{x,t+1}^{*}\eta_{x,t})\Big]
\notag\\ 
&-\sum_x\big[ \psi^{*}_{x,t+1} (\psi_{x,t+1}-\psi_{x,t}  )+\eta^{*}_{x,t+1} (\eta_{x,t+1}-\eta_{x,t}) \big] \Big\}~.\label{eq:partfunc2}
\end{align}
Here we have introduced the ``charge field''
\be
Q_{x,t,t'}=\psi_{x,t}^{*}\psi_{x,t'}-\eta_{x,t}^{*}\eta_{x,t'}~.
\ee

To simulate Eq.~(\ref{eq:partfunc2}) via standard
Monte-Carlo methods one wishes to eliminate
the Grassmann variables and ultimately deal only with regular
complex numbers. The customary way to achieve this which
is applied to various systems with fermionic degrees of freedom,
is to integrate out the fermion fields and to rewrite the fermionic
part of the action as a determinant, which
then can be sampled stochastically
using pseudo-fermion sources.
To this end a Gaussian integral of the form
\be
\int \left[\prod_{m=1}^n\,d\chi_m^{*}\,d\chi_m
\right] \,e^{-\sum\limits_{i,j=1}^{n}\chi_i^{*}H_{ij}\chi_j}
\propto [\det H ]^{\pm 1}~,\label{eq:gaussint1}
\ee
must be carried out (this identity holds for both complex
commuting variables $\chi_i$ with the negative sign on the right-hand
side, in case of which $H$ must have a positive Hermitian part, and
for Grassmann variables with the positive sign and no restrictions
on $H$). This is impossible for the current form of Eq.~(\ref{eq:partfunc2}) 
since forth powers of the Grassmann (``field'') variables appear.
We can eliminate these at the expense of introducing a scalar
auxiliary field $\phi$ by applying the Hubbard-Stratonovich
transformation,

\begin{align}
\exp\Big\{ -\frac{\delta}{2} \sum_{t=0}^{N_t-1}  
\sum_{x,y}Q_{x,t+1,t}V_{xy}Q_{y,t+1,t} \Big\}
\propto \int \mathcal{D}\phi\, \exp\Big\{&-\frac{\delta}{2} \sum_{t=0}^{N_t-1} \sum_{x,y} \phi_{x,t} V_{xy}^{-1} \phi_{y,t}\notag \\
&-i\,\delta
\sum_{t=0}^{N_t-1} \sum_{x} \phi_{x,t} Q_{x,t+1,t}\Big\}~. \label{eq:hubbard1}
\end{align}
We will refer to $\phi$ as the Hubbard-Coulomb field. It has the
dimension of energy or inverse length as usual.
A constant factor in Eq.~(\ref{eq:hubbard1})
is omitted, since it can be absorbed into the measure. 
Here we have introduced the shorthand notation
\begin{align}\mathcal{D}\phi=\left[  \prod_{t=0}^{N_t-1} \prod_x
d\phi_{x,t} \right]~,
\end{align}
which we will use for the remainder of the paper for any generic field
$\chi$.
Applying the transformation, we obtain
\begin{align}
\Tr e^{-\beta H} &= \int \mathcal{D}\psi\, \mathcal{D}\psi^*\, 
\mathcal{D}\eta\, \mathcal{D}\eta^*\, \mathcal{D}\phi\
\exp \Big\{ -\delta \sum_{t=0}^{N_t-1} \Big[\frac{1}{2}\sum_{x,y} \phi_{x,t}V_{xy}^{-1} \phi_{y,t} + \sum_x  i \phi_{x,t} Q_{x,t+1,t} \notag\\
&-\sum_{\langle x,y \rangle}\kappa
(\psi_{x,t+1}^{*} \psi_{y,t}+\psi_{y,t+1}^{*} \psi_{x,t}+\eta_{y,t+1}^{*} \eta_{x,t}+\eta_{x,t+1}^{*} \eta_{y,t}  )
\notag\\ &+\sum_x m_s(\psi_{x,t+1}^{*}\psi_{x,t}+\eta_{x,t+1}^{*}\eta_{x,t})
+\frac{1}{2} \sum_x V_{xx}(\psi_{x,t+1}^{*}\psi_{x,t}+\eta_{x,t+1}^{*}\eta_{x,t})\Big]\notag\\ 
&-\sum_{t=0}^{N_t-1} \sum_x\big[ \psi^{*}_{x,t+1} (\psi_{x,t+1}-\psi_{x,t}  )+\eta^{*}_{x,t+1} (\eta_{x,t+1}-\eta_{x,t}) \big] \Big\}~.
\label{eq:partfunc3}
\end{align}
This expression contains no fourth powers.
By introducing a matrix $M$, which is defined in terms of its elements as
\begin{align}
M_{(x,t)(y,t')}~=~& \delta_{xy}(\delta_{tt'}-\delta_{t-1,t'})-
\kappa\frac{\beta}{N_t} \sum\limits_{\vec{n}} \delta_{y,x+\vec{n}}\delta_{t-1,t'}+ 
m_s\frac{\beta}{N_t} \delta_{xy} \delta_{t-1,t'} \notag \\
&+\frac{V_{xx}}{2} \frac{\beta}{N_t}\delta_{xy}\delta_{t-1,t'}
+i\phi_{x,t}\frac{\beta}{N_t}\delta_{xy} \delta_{t-1,t'}~.\label{eq:fermionmatrix}
\end{align}
we can rewrite Eq.~(\ref{eq:partfunc3}) as
\begin{align}
\Tr e^{-\beta H}=&\int \mathcal{D}\psi\, \mathcal{D}\psi^*\, 
\mathcal{D}\eta\, \mathcal{D}\eta^*\, \mathcal{D}\phi\
\exp\bigg\{-\frac{\delta}{2}\sum_{t=0}^{N_t-1}\sum_{x,y} \phi_{x,t}V_{xy}^{-1} \phi_{y,t} 
\notag\\
&-\sum_{t,t'=0}^{N_t-1}\sum_{x,x'}
\Big[ \psi^{*}_{x,t} M_{(x,t)(x',t')} \psi_{x',t'}+
 \eta^{*}_{x,t} M^{*}_{(x,t)(x',t')} \eta_{x',t'} \Big]
\bigg\}
~.\label{eq:partfunc4}
\end{align}
Here $M^*$ means complex conjugate (of individual elements), 
not Hermitian conjugate (which we will write as $M^\dagger$).
The notation $(x,t)$ here is understood to imply that
the indices labeling matrix elements of $M$ enumerate pairs of
coordinates $x$ and $t$.

We can now carry out the Gaussian integration and obtain
\begin{align}
\Tr e^{-\beta H}&=\int \mathcal{D}\phi
\, \det M^*(\phi) \det M(\phi)
\exp \left\{-\frac{\delta}{2} \sum_{t=0}^{N_t-1} \sum_{x,y}
\phi_{x,t}V_{xy}^{-1} \phi_{y,t}
\right\}\notag\\
&=\int \mathcal{D}\phi
\, \det\left[ M(\phi) M^\dagger(\phi)  \right]
\exp \left\{-\frac{\delta}{2} \sum_{t=0}^{N_t-1} \sum_{x,y}
\phi_{x,t}V_{xy}^{-1} \phi_{y,t} \right\}\label{eq:partfunc5}
\end{align}
Here we can see explicitly why it was convenient
to introduce hole operators for the
spin-down states. In absence of magnetic fields and spin
dependent interactions, the true spin of the electrons formally plays the
role of a flavor quantum number. The corresponding flavor symmetry
entails that, without mass term, both spin degrees of freedom lead to
the same fermion matrix. By introducing hole operators for one species
we basically changed the sign of the charge of the corresponding
quasiparticles and hence obtain a manifestly real and positive
contribution  $\det( M M^\dagger )$ in the measure for the product of
both flavors. Note, however, that because $\det M^*(m_s) = \det M(-m_s)$,
it is only the mass term which can lead to truly complex fermion
determinants.  Without it, $M$ would have definite
Hermiticity properties for each spin. The decisive assumption for a
real product of fermion matrices here is the spin staggering of the
mass term. Without introduction of hole operators for one of the spin
states, both fermionic quasiparticles would have the same charge, and
hence the same sign in the last term of Eq.~(\ref{eq:fermionmatrix}),
but the opposite sign in the mass term proportional to $m_s$. Only if
that were not the case, would the product of fermion determinats not
be real.

Eq.~(\ref{eq:partfunc5}) is now in a form suitable for simulation
via Hybrid-Monte-Carlo. Note that the only remaining integration
is over the Hubbard-Coulomb field $\phi$. This field will ultimately
be the only dynamical field which must be stored in computer memory
and will represent the ``lattice configuration'' of the system. 

One final remark must be made: Simulations based
on Eq.~(\ref{eq:fermionmatrix}) suffer
from a severe technical problem, due to the fact that
$\phi$ is a non-compact field.
While the expression is formally correct, the determinant
implicitly contains powers of $\phi$ up to $\phi^N$, where $N$ is the
number of lattice sites. This causes numerical instabilities due to
an uncontrollable amplification of rounding errors
(in fact, we have previously verified this explicitly \cite{Smith:2013pxa}).
A solution to this problem was also worked out in 
Ref.~\cite{Brower:2012zd} already, where it was
shown that one can make the replacement
\be
-\delta_{xy}\delta_{t-1,t'}+
\frac{\beta}{N_t} \frac{V_{xx}}{2}
\delta_{xy}\delta_{t-1,t'}+i\frac{\beta}{N_t}\phi_{x,t}\delta_{xy} \delta_{t-1,t'}
\longrightarrow 
-e^{-i\frac{\beta}{N_t}\phi_{x,t}} \delta_{xy} \delta_{t-1,t'} ~,\label{eq:subcompact}
\ee
which introduces a compact field that is numerically stable
(the determinant then contains sums over $\phi$ rather than
products). The argument presented in Ref.~\cite{Brower:2012zd}
is based on the observation that the fermionic part of the
Hubbard-Stratonovich transformed action corresponding to the path
integral in Eq.~(\ref{eq:partfunc4}) in the temporal continuum limit
$\delta\to 0$ becomes invariant under spatially constant
one-dimensional (temporal) gauge transformations. 
This gauge invariance is maintained in the discretized version when
the compact field is used as a parallel transporter between time
slices in the fermion matrix. The two discretizations are equivalent
in the continuum limit. An alternative way
to see this is to apply the Hubbard-Stratonovich transformation
directly on the operator level rather than the field level, i.e.~in the form
\begin{align}
\exp\left\{ -
\frac{\delta}{2}\sum_{x,y}q_{x}V_{xy}q_{y} \right\}
\propto \int \left[ \prod_x \phi_x \right]\, \exp\left\{-\frac{\delta}{2}
\sum_{x,y} \phi_{x} V_{xy}^{-1} \phi_{y}
-i\,\delta
\sum_{x} \phi_{x} q_{x}\right\}~, \label{eq:hubbard2}
\end{align}
already in Eq.~(\ref{eq:partfunc}). Rather than
using Eq.~(\ref{eq:coher_theorem1}) for the normal ordered form of the
interaction on the left hand side, one then computes
the matrix elements of the last term on the right 
containing the charge operator
$q_x$ using that \cite{Ulybyshev:2013swa} 
\be
\langle \xi |\, e^{\sum_{x,y}\axd A_{xy} \ay }\, |\xi' \rangle = 
\exp\left( \sum_{x,y} \xi^*_x \left( e^A \right)_{xy} \xi'_y \right)~,
\label{eq:matrixelement2}
\ee
which holds for arbitrary matrices $A$. Here, $A$ is a diagonal matrix with  
elements $A_{xx}=\pm i \delta\, \phi_x$. One then obtains the
fermion operator directly as
\begin{align}
M_{(x,t)(y,t')}~=~& 
\delta_{xy}(\delta_{tt'}-e^{-i\frac{\beta}{N_t}\phi_{x,t}} \delta_{t-1,t'})-
\kappa\frac{\beta}{N_t} \sum\limits_{\vec{n}} \delta_{y,x+\vec{n}}\delta_{t-1,t'}+ 
m_s\frac{\beta}{N_t} \delta_{xy} \delta_{t-1,t'}~,
\end{align}
in which $\exp(-i\frac{\beta}{N_t}\phi_{x,t})$ acts as the parallel transporter
in the Euclidean time direction.

\subsection{The Hybrid-Monte-Carlo algorithm}
Our principal objective is to compute expressions of the form
\be
\langle O \rangle = \int \mathcal{D}\phi\, O(\phi) P(\phi)~,
\ee
where
\be
P(\phi)= \frac{1}{Z}\det\left[ M(\phi) M^\dagger(\phi)  \right]
e^{-S(\phi)}~.\label{eq:fieldweight1}
\ee
Here $\phi$ is the Hubbard-Coulomb field introduced in the preceding
paragraph,\footnote{Note 
that we have suppressed the indices here, i.e. $\phi_{x,t}\equiv\phi$.
 $\phi$ is then understood as denoting a vector whose components are
labeled by space-time indices. Where appropriate, we use such short-hand
notation in the following for the Hubbard-Coulomb field
as well as the pseudofermion fields. Analogously, we use a vector/matrix
notation for quadratic forms such as $\sum_{i,j} \chi^*_i A_{ij}
\eta_j \equiv \chi^\dagger A \eta$ where $i,j$ label all pairs $(x,t)$. }
$O(\phi)$ is some function of this field which represents
an observable, $S(\phi)$ is the part
of the action which depends on $\phi$ only, and $\det( M M^\dagger )$ 
accounts for the effects of the fermions. This problem is precisely
of the form which is encountered in lattice gauge theory,
where one has fermionic degrees of freedom (usually quark fields) 
interacting via gauge fields. Here we only have the scalar field $\phi$
which represents the partially screened Coulomb field as the
zero component of the Abelian photon field in the instantaneous
approximation.  

The Monte-Carlo approach to the problem is
to generate representative configurations of the field $\phi$ with
probability $P(\phi)$ and measure observables $O(\phi)$ 
on these configurations. 
A well established algorithm to generate these configurations
is the Hybrid-Monte-Carlo algorithm \cite{Duane:1987de, Clark:2006fx,
  Gattringer:2010}. 
The first step is to represent the fermion determinant
as a Gaussian integral over pseudofermion fields 
$\chi$ (these are commuting complex field variables rather
than Grassmann numbers) using Eq.~(\ref{eq:gaussint1}) with
the negative sign on the right-hand
side. We thus obtain
\begin{align}
\Tr e^{-\beta H}=&\int \mathcal{D}\phi\,
\mathcal{D}\chi\,\mathcal{D}\chi^*\,
\exp\Big\{-\frac{\delta}{2}\sum_{t=0}^{N_t-1}\sum_{x,y} \phi_{x,t}V_{xy}^{-1} \phi_{y,t} 
-\sum_{t,t'=0}^{N_t-1}\sum_{x,x'}
\chi^{*}_{x,t}( M M^{\dagger} )^{-1}_{(x,t)(x',t')} \chi_{x',t'}
\Big\}
~.
\end{align}
Note that the inverse of $M M^{\dagger}$ now appears in the
exponent.\footnote{Since  
$M M^{\dagger}$ is a positive Hermitian matrix, given a complex
source vector $A$ one can efficiently compute
$B=( M M^{\dagger} )^{-1} A$ using the conjugate gradient algorithm.
For an excellent (unpublished) pedagogical presentation see
\emph{An Introduction to the Conjugate Gradient Method
Without the Agonizing Pain} by J.~Shewchuk which
can be found on the website of the author.}
 
Now that we have introduced the pseudofermions, we are faced with
the problem of generating $\phi,\chi$ according to a joint
probability density
\be
P(\phi,\chi)= \frac{1}{Z} e^{-S(\phi)-S'(\chi) }~.\label{eq:fieldweight2}
\ee
The way to approach this problem is to generate $\phi$ and $\chi$ separately:
One combines a heatbath update of the pseudofermions
(the distribution of $\chi$ thus is generated ``directly''
as $P(\chi)\sim e^{-\chi^\dagger(M M^\dagger)^{-1} \chi}$) with a molecular
dynamics (MD) evolution of the Hubbard-Coulomb field. 
The MD evolution is
a fictitious dynamical process which evolves the field $\phi$
in computer time $\tau$ from some starting configuration in such a way that,
after a suitable thermalization time, the propagation through phase
space is consistent with the distribution defined by Eq.~(\ref{eq:fieldweight2}). It is generated by introducing a fictitious 
momentum field $\pi$ associated with the $\tau$-evolution of $\phi$.
This then defines a Hamiltonian
$\calh = S(\phi)+S'(\chi)+\frac{\pi^2}{2}$ (which implies that $\pi$ is distributed
as $P(\pi)\sim e^{-\pi^2/2}$).
The Hubbard field $\phi$ is allowed to evolve by
integrating Hamilton's equations for $\phi$ and $\pi$. 

Since the equations cannot be solved exactly a symplectic integrator
such as the Leapfrog integrator is used (discussed below), which introduces an
error (difference in energy $\Delta\calh$) 
from the finite step-sizes used in such integrators. This is then corrected
by performing a Metropolis check at the end of each trajectory.
The complete HMC algorithm thus consists of repeating the following steps:

\begin{samepage}
\begin{itemize}
\item{Update the momentum field $\pi$ using Gaussian noise: $P(\pi)\sim e^{-\pi^2/2}$}~.
\item{Update pseudofermions $\chi$ by generating another complex 
field $\rho$ with
$P(\rho)=e^{-\rho^\dagger \rho}$ and then obtaining $\chi=M \rho$~.}
\item{Generate a molecular dynamics trajectory.}
\item{Perform Metropolis check to correct step-size error: Accept new configuration with
probability $P=\min(1,e^{-\Delta \calh})~.$}
\end{itemize}
\end{samepage}

For the reminder of this paragraph we will
discuss the details of the MD evolution which
is the central part of the algorithm. Consider
that we have introduced the momentum field $\pi$. We thus
have the Hamiltonian
\be
\calh=\frac{\delta}{2}\sum_{t=0}^{N_t-1}\sum_{x,y} \phi_{x,t}V_{xy}^{-1} \phi_{y,t} 
+\sum_{t,t'=0}^{N_t-1}\sum_{x,x'}
\chi^{*}_{x,t}( M M^{\dagger} )^{-1}_{(x,t)(x',t')} \chi_{x',t'}
+\frac{1}{2} 
\sum_{t=0}^{N_t-1}\sum_{x} \pi_{x,t}^2 \, .
\ee
In vector/matrix notation this assumes the compact form:
\be
\calh=\frac{\delta}{2} \phi^T V^{-1}\phi + \chi^\dagger (M
M^{\dagger})^{-1} \chi +\frac{\pi^T \pi}{2} \, .
\ee
Hamilton's equation are then given by
\be
\left[ \frac{d \phi}{d \tau} \right]^T= \frac{\partial \calh}{\partial \pi}~,
\quad
\left[ \frac{d \pi}{d \tau} \right]^T= - \frac{\partial \calh}{\partial \phi}~.
\ee
Given a set of initial conditions $(\phi,\pi)$, solutions to this
set of equations can be approximated numerically using Leapfrog
integration: Assume
that the time derivatives are approximated by finite differences
\be
\frac{d \phi}{d \tau} \approx \frac{1}{\epsilon}[\phi(\tau+\epsilon)- \phi(\tau)]
\equiv\frac{1}{\epsilon}(\phi_{\tau+1}- \phi_{\tau})~,\quad
\frac{d \pi}{d \tau} \approx \frac{1}{\epsilon}[\pi(\tau+\epsilon)- \pi(\tau)]
\equiv\frac{1}{\epsilon}(\pi_{\tau+1}- \pi_{\tau})~.
\ee
We then define position steps $V_{\phi}(\epsilon)$ and momentum steps $V_{\pi}(\epsilon)$ as
\begin{align}
V_{\phi}(\epsilon)~:& \quad \phi_{\tau+1}= \phi_\tau+ \epsilon (\partial \calh /\partial \pi )^T~,\notag\\
V_{\pi}(\epsilon)~:& \quad \pi_{\tau+1}= \pi_\tau- \epsilon (\partial \calh /\partial \phi )^T~.
\end{align}
Leapfrog integration consists of iterating combinations of steps of
the form
\begin{align}
V_{\pi}(\epsilon/2)V_{\phi}(\epsilon)V_{\pi}(\epsilon/2)
\quad \textrm{or} \quad
V_{\phi}(\epsilon/2)V_{\pi}(\epsilon)V_{\phi}(\epsilon/2)
\label{eq:leapfrog1}
\end{align}
until a desired trajectory length $L=N \epsilon$ is reached.
The former is known as PQP integration, the latter as QPQ integration.
Leapfrog integration does not conserve energy. The error
(deviation from the ``true'' trajectory through phase-space)
can be quantified by the difference in energy $\Delta\calh$
and is of order $\Delta\calh \sim \calo (\epsilon^2)$.

To derive the expressions for the derivatives of $\calh$ we use matrix calculus.
We assume a numerator convention, i.e.~vectors $\rho$, $\sigma$ which
are defined as columns, and derivatives which are defined as
\begin{align}
&\frac{d \sigma}{d \rho}=
\begin{pmatrix}
\frac{\partial \sigma_1}{\partial \rho_1}&\cdots & \frac{\partial \sigma_1}{\partial \rho_n}\\
\vdots&\ddots &\vdots\\
\frac{\partial \sigma_m}{\partial \rho_1}&\cdots & \frac{\partial \sigma_m}{\partial \rho_m}
\end{pmatrix}
,~\quad
\frac{d M}{d s }=
\begin{pmatrix}
\frac{\partial M_{11}}{\partial s}&\cdots & \frac{\partial M_{1m}}{\partial s}\\
\vdots&\ddots &\vdots\\
\frac{\partial M_{n1}}{\partial s}&\cdots & \frac{\partial M_{nm}}{\partial s}
\end{pmatrix}
,~\quad \frac{d \sigma}{d s}=
\begin{pmatrix}
\frac{\partial \sigma_1}{\partial s}\\
\vdots\\
\frac{\partial \sigma_m}{\partial s}
\end{pmatrix}
,~\notag\\
&\frac{d s}{d \rho}=\left(\frac{\partial s}{\partial \rho_1},\ldots, \frac{\partial s}{\partial \rho_n} \right),
\end{align}
where $s$ is a scalar and $M$ is a matrix. The following identities
then hold:
\begin{align}
&\frac{d( \sigma^T \sigma) }{d\rho}=2\,\sigma^T \frac{d \sigma }{d\rho}
~,~\quad \frac{d( \rho^T A \rho) }{d\rho} =2\, \rho^T A 
~,~\quad \frac{d U^{-1} }{d s}= -U^{-1}\frac{d U}{d s}  U^{-1}~,~\notag\\
&\frac{d(U V)}{d s}= U \frac{d V}{d s}+\frac{d U}{d s} V~.
\end{align}
Here $U,V$ are matrices which depend on $s$, and $A$ is a symmetric
matrix which does not depend on $\rho$.

It is now easy to see that
\be
\left[ \frac{d \phi}{d\tau} \right]^T= \frac{\partial \calh}{\partial \pi}= 
\pi^T
\ee
The second equation requires a little more work. It consists of
two terms, which are the force terms generated by the Hubbard-Coulomb 
and the fermion fields. Using the above identities, it follows that
\be
\left[ \frac{d \pi}{d\tau} \right]^T= - \frac{\partial \calh}{\partial \phi}
=- {\delta}\, \phi^T V^{-1} - \frac{\partial }{\partial \phi}
\left[\chi^\dagger (M M^{\dagger})^{-1} \chi \right] \equiv
F_\phi + F_\chi~ .
\ee
Deriving the Hubbard force $F_\phi$ was straightforward. 
To obtain the the fermion force $F_\chi$ we
note that it is vector valued. It can be evaluated component-wise
as
\begin{align}
(F_\chi)_k&=-\frac{\partial }{\partial \phi_k} \left[
\chi^\dagger (M M^{\dagger})^{-1} \chi\right]
\notag\\
&=-\chi^\dagger \left[ \frac{\partial (M M^{\dagger})^{-1} }{\partial \phi_k}
 \right] \chi
= \chi^\dagger (M M^{\dagger})^{-1} \left[ \frac{\partial (M M^{\dagger})  }{\partial \phi_k}
\right](M M^{\dagger})^{-1} \chi\notag\\
&= \chi^\dagger (M M^{\dagger})^{-1}  \left[
 \frac{\partial M }{\partial \phi_k}  M^{\dagger}
+ M\frac{\partial M^\dagger }{\partial \phi_k} \right](M M^{\dagger})^{-1} \chi
= \eta^\dagger \frac{\partial M }{\partial \phi_k} \xi
+\xi^\dagger \frac{\partial M^\dagger }{\partial \phi_k}\eta\notag\\
&=2\, \textrm{Re}\left[\eta^\dagger \frac{\partial M }{\partial \phi_k} \xi \right]~.
\end{align}
Here we have introduced the notation 
\begin{align}
\eta = (M M^{\dagger})^{-1}\chi~,\quad \xi = M^{-1}\chi = M^{\dagger} \eta~.
\end{align}
The precise form of the fermion force now depends on the choice of $M$.
For the fermion-matrix defined in Eq.~(\ref{eq:fermionmatrix}) we obtain
\begin{align}
(F_\chi)_{(x,t)}=-\frac{\partial }{\partial \phi_{(x,t)}}& \left[
\chi^\dagger (M M^{\dagger})^{-1} \chi\right]
=-2 \frac{\beta}{N_t} \textrm{Im}\left[ \eta^*_{(x,t)} \xi_{(x,t-1)} \right]
\end{align}
For the version which uses the compact Hubbard field
(with the substitution as shown in Eq.~(\ref{eq:subcompact}))
we obtain
\begin{align}
(F_\chi)_{(x,t)}=-\frac{\partial }{\partial \phi_{(x,t)}} \left[
\chi^\dagger (M M^{\dagger})^{-1} \chi\right]
=-2 \frac{\beta}{N_t} \textrm{Im}\left[ \eta^*_{(x,t)} 
e^{-i\frac{\beta}{N_t}\phi_{x,t} }\xi_{(x,t-1)} \right]
\label{eq:fermforce_compact}
\end{align}
Lastly, it should be pointed out that it is often possible
to choose a numerical integration scheme which performs
better than the standard Leapfrog integration
defined in Eq.~(\ref{eq:leapfrog1}). In particular, when the force
$F=d \pi / d\tau$ is, as is the case here, composed of
different components $F_i$ which differ both in magnitude
and associated computational cost, one may obtain a more efficient
integrator by decomposing the momentum steps $V_{\pi}(\epsilon)$
into sub-steps $V_{\pi}^{F_i}(\epsilon)$ which each use only one of the
force components. If, for example, one has $F=F_1+F_2$, where
$F_2$ is both much cheaper to compute and of larger magnitude than
$F_1$, one may gain performance by using the decomposition
\be
V_{\pi}(\epsilon)\quad\to\quad V_{\pi}^{F_1}(\epsilon/2)
\left[V_{\pi}^{F_2}(\epsilon/m) \right]^m
V_{\pi}^{F_1}(\epsilon/2)\label{eq:swleapfrog}
\ee
This is known as Sexton-Weingarten integration \cite{Sexton:1992}.
The constant $m$ must be tuned to the particular problem. It is
often (but not always) a good idea to tune $m$ such that
the force components entering into Eq.~(\ref{eq:swleapfrog})
are of a similar magnitude. 

\subsection{The order-parameter}

We wish to investigate spontaneous breaking of sublattice
symmetry. Thus we require a proper order parameter as determined by
the mass term which acts as the explicitly symmerty breaking external
field. The corresponding choice here is to use the difference of the
spin density operators on the two sublattices $A$ and $B$,
\be
\Delta_N =n_A-n_B=\frac{1}{L_x L_y}\left\{\sum\limits_{x\in X_A}( a^\dagger_{x} a_{x} + b^\dagger_{x} b_{x} )
-\sum\limits_{x \in X_B}( a^\dagger_{x} a_{x} + b^\dagger_{x} b_{x} )\right\}.
\ee
Its expectation value is given by
\be
\langle \Delta_N \rangle = \frac{1}{Z} \Tr\left[ \Delta_N e^{-\beta H} \right]~,
\ee
which in the functional integral form is expressed as 
\begin{align}
\langle \Delta_N \rangle =&\frac{1}{ZN_tL_x L_y}\int
\mathcal{D}\psi\,\mathcal{D}\psi^*\,\mathcal{D}\eta\,\mathcal{D}\eta^*
\notag\\
&\quad\quad\quad\times
\Big\{ \sum_{X_A,t}\left(\psi^*_{x,t+1}\psi_{x,t}+
\eta^*_{x,t+1}\eta_{x,t}\right)- 
\sum_{X_B,t}\left(\psi^*_{x,t+1}\psi_{x,t}+\eta^*_{x,t+1}\eta_{x,t}\right) 
\Big\}e^{-\beta H} \notag\\
=&\frac{-1}{\beta ZL_x L_y} \frac{\partial}{\partial m} 
\int\mathcal{D}\psi\,\mathcal{D}\psi^*\,\mathcal{D}\eta\,\mathcal{D}\eta^*
\,e^{-\beta H}
=\frac{-1}{\beta ZL_x L_y}\int  \mathcal{D}\phi
\left[ \frac{\partial}{\partial m} \det\left(M M^\dagger \right)
 \right] e^{-S[\phi]}\notag\\
=&\frac{-1}{\beta ZL_x L_y}\int\mathcal{D}\phi\,
\det \left(M M^\dagger \right)
\Tr\left[ M^{-1}
\frac{\partial \left(M M^\dagger \right)}{\partial m}
{M^{-1}}^\dagger  \right] e^{-S[\phi]} 
\notag\\
=&\frac{-2}{\beta ZL_x L_y}\int\mathcal{D}\phi\,
\det \left(M M^\dagger \right)
\ReTr\left[ M^{-1} \frac{\partial M}{\partial m}   \right] e^{-S[\phi]} 
\end{align}
As the magnetization in a classical spin system, it is of course given by the  
derivative with respect to the external field. 
Using Eq.~(\ref{eq:fermionmatrix}) we then obtain
\begin{align}
\langle \Delta_N \rangle=\frac{-2}{N_tL_x L_y} \sum\limits_{t=0}^{N_t-1}
\Re\left\langle \sum\limits_{x \in X_A} M^{-1}_{(x,t)(x,t+1)}-
\sum\limits_{x \in X_B} M^{-1}_{(x,t)(x,t+1)} \right\rangle ~.\label{eq:pbp1}
\end{align}
This expression is very similar to the lattice formulation of the 
chiral condensate in QCD.
It holds for both, compact and non-compact Hubbard-Coulomb fields.

Computing this order parameter hence amounts to computing the expectation
value of a trace of an operator, i.e. 
\be
\langle \Delta_N \rangle =\Re \langle \Tr\left[ D^{-1} \right] \rangle~,
\ee
with $D$ defined appropriately. Straightforward
calculation of such a trace is not feasible. A widely used method to deal with 
this problem is the noisy estimator approach
(see e.g.~Ref.~\cite{Gattringer:2010}). Using Gaussian noise vectors, i.e.
complex pseudofermion sources $\chi$ randomly drawn from 
$P(\chi)\sim \exp(-\chi^\dagger \chi)$, one can estimate the trace
(on a given lattice configuration)
as
\be
\Tr\left[ D^{-1} \right] \approx \frac{1}{K} 
\sum\limits_{k=1}^K {\chi^{(k)}}^\dagger D^{-1} \chi^{(k)}
\ee
Here $K$ is the total number or source vectors and $k$ is the index
which labels them. 
The accuracy of the estimate becomes successively better with increasing $K$.

\subsection{Second order fermion operator}

\begin{figure}
\begin{center}
\resizebox{0.49\textwidth}{!}{%
\includegraphics{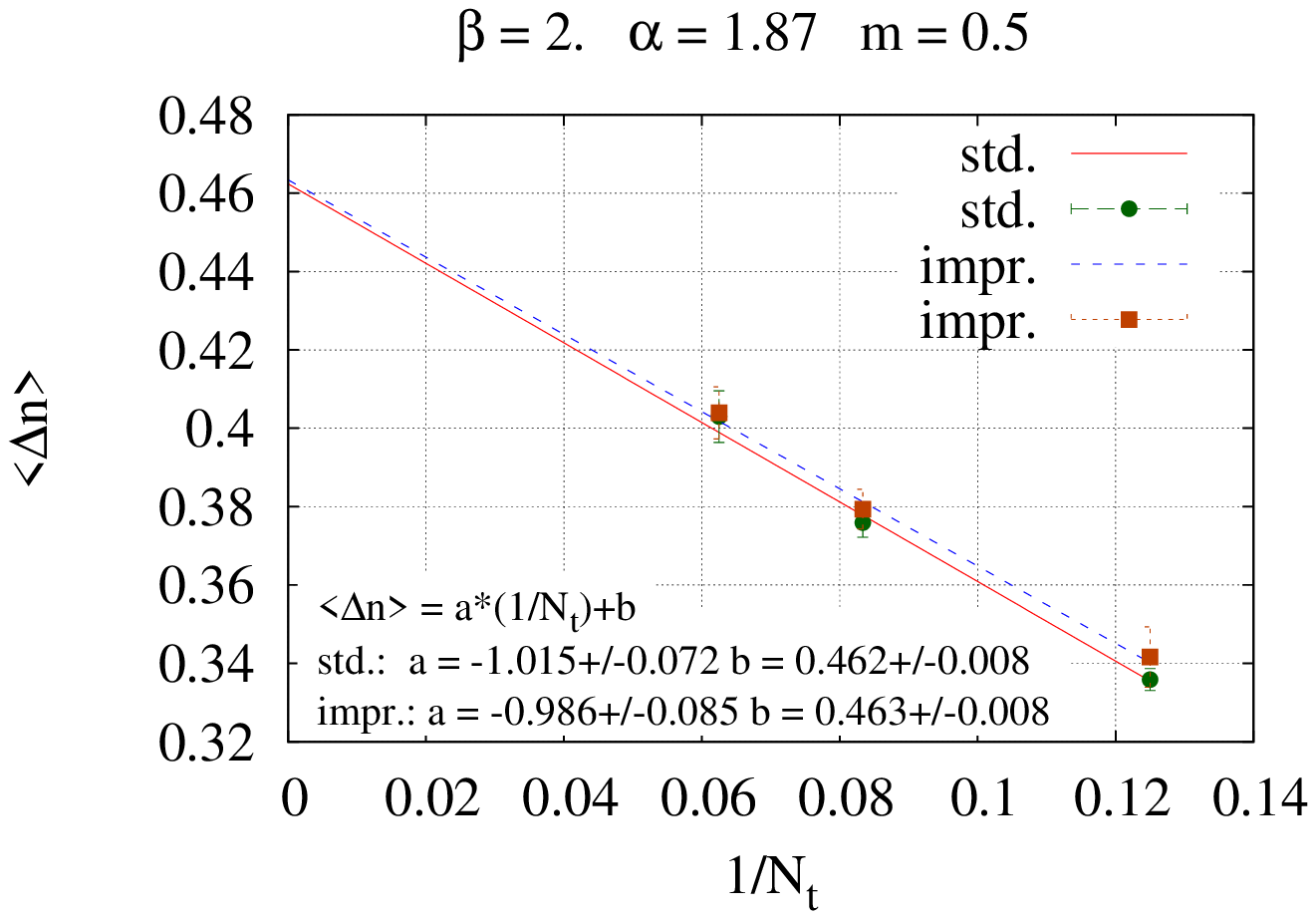}}
\resizebox{0.49\textwidth}{!}{%
\includegraphics{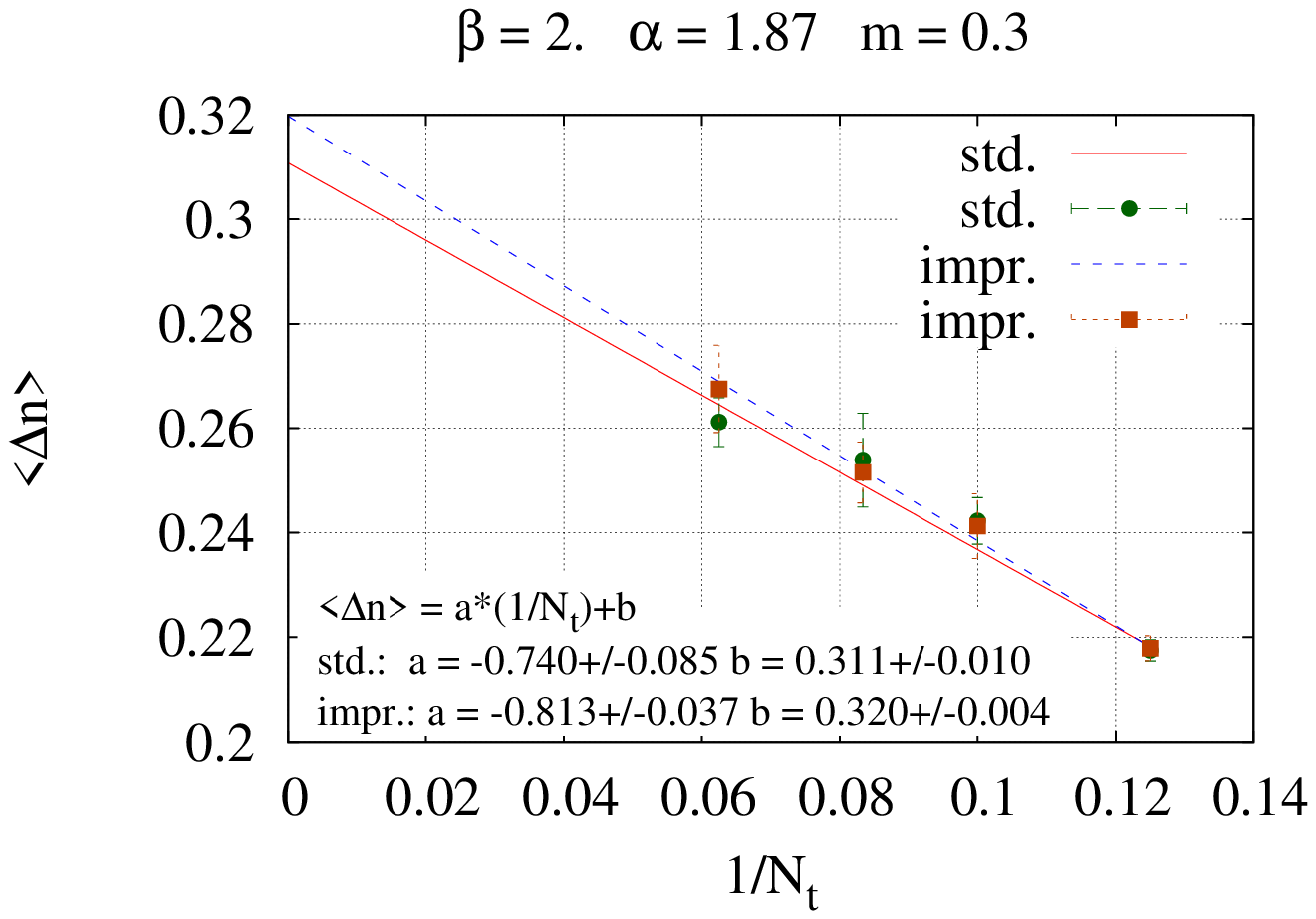}}
\end{center}
\caption{\label{fig:compare_discr} Comparison of $1/N_t$ scaling of
$\langle\Delta_N\rangle$ on $L_x=L_y=6$ lattices. The first and second order
discretizations are referred to as ``std.'' and ``impr.'' respectively
in the figures. The potential chosen for these simulations
includes the effect of mirror charges and is discussed in Ref. \cite{Smith:2013pxa}. }
\end{figure}

Eq.~(\ref{eq:fermionmatrix}) is by far not the only possible
form for the fermion matrix. As was discussed in Ref.~\cite{Brower:2012zd},
there is a great amount of freedom in discretizing the Euclidean
time direction which could, in principle, be exploited to construct improved
actions that approach the continuum limit faster. 
A particular second order discretization scheme was proposed in
Ref.~\cite{Ulybyshev:2013swa}. We have previously
obtained some evidence (on small lattices and with
a potential which differs from the one used in this work) 
that this version is,
in terms of discretization errors affecting
the order-parameter, equivalent to the direct discretization
discussed in the preceding paragraph
and doesn't yield any improvement. 
In Fig. \ref{fig:compare_discr} we show two examples.
See Ref.~\cite{Smith:2013pxa}
for further details.
An extensive auto-correlation
analysis has not yet been done, but would be useful
to determine whether there is any performance gain.
In any case, for the purpose of cross-checking our results we have 
implemented this version in our code also. 
We sketch the derivation in this paragraph. In particular
we highlight which aspects of the preceding discussions
are changed.

Consider that we have introduced coherent states 
and derived Eq.~(\ref{eq:partfunc}). At this point,
one may choose to factor the exponential in such a way, that
the interacting part is split off,
\be
e^{-\delta H}=e^{-\delta(H_c + H_{tb}+H_m)}~ \to~ e^{-\delta(H_{tb}+H_m)}\,e^{-\delta H_c}~.
\ee
We insert an additional set of coherent states between the 
factors and relabel all states to restore the lexicographic ordering
of their index (in fact, compared to previously, the enumeration of the 
t-coordinate is reversed in order to match the conventions of
Ref.~\cite{Ulybyshev:2013swa}, but this 
is of no further relevance since it is a trivial relabeling).
We then obtain
\begin{align}  
\Tr e^{-\beta H}=&\int \left[\prod_{t=0}^{2N_t-1} \prod_x
d\psi^{*}_{x,t} \, d\psi_{x,t} \, d\eta^{*}_{x,t} \, d\eta_{x,t} \right] \notag\\
&\times\prod_{t=0}^{N_t-1}~\exp\left\{-\sum_x(  \psi^{*}_{x,2t} \psi_{x,2t}+\eta^{*}_{x,2t} \eta_{x,2t}
+\psi^{*}_{x,2t+1} \psi_{x,2t+1}+\eta^{*}_{x,2t+1} \eta_{x,2t+1} ) \right\}\notag\\
&\times \langle \psi_{2t},\eta_{2t}|e^{-\delta (H_{tb}+H_m)}|\psi_{2t+1},
\eta_{2t+1}\rangle \langle \psi_{2t+1},\eta_{2t+1}|e^{-\delta H_c}|
\psi_{2t+2},\eta_{2t+2}\rangle~.
\end{align}
Computing the matrix elements and introducing the compact Hubbard-Coulomb field
as previously now yields
\begin{align}  
\Tr e^{-\beta H}=&\int \left[\prod_{t=0}^{2N_t-1} \prod_x
d\psi^{*}_{x,t} \, d\psi_{x,t} \, d\eta^{*}_{x,t} \, d\eta_{x,t} \right]
\left[\prod_{t=0}^{N_t-1} \prod_x d\phi_{x,t} \right]
e^{-(S_\psi + S_\eta)}\notag\\
&\times \exp\left\{- \frac{\delta}{2}\sum_{t=0}^{N_t-1}\sum_{x,y} \phi_{x,t}V_{xy}^{-1} \phi_{y,t}
\right\}~,
\end{align}
where we have
\begin{align}
S_\psi& = \sum\limits_{t=0}^{N_t-1}
 \left[ \sum\limits_x \psi^{*}_{x,2t} \left(\psi_{x,2t}-\psi_{x,2t+1}\right)  \right.
 -\delta \, \kappa \sum\limits_{<x,y>}
 \left( \psi^{*}_{x,2t} \psi_{y,2t+1} + \psi^{*}_{y,2t} \psi_{x,2t+1} \right)
 \notag \\
& + \sum\limits_x \psi^{*}_{x,2t+1}
\left(\psi_{x,2t+1}  - e^{-i\delta \, \phi_{x,t}} \psi_{x,2t+2} \right)
 +  \left. \delta \sum\limits_x m_s \psi^{*}_{x,2t} \psi_{x,2t+1} \right]~,
\end{align}
and an analogous expression for $S_\eta$ with the opposite charge,
i.e.~with a phase factor $e^{i\delta\, \phi_{x,t}}$. 
By introducing a fermion matrix of the form
\be
M_{(x,t)(y,t')}  = \left\{
\begin{array}{lr}
\delta_{xy}(\delta_{tt'}-\delta_{t+1,t'})-\frac{\beta}{N_t} \kappa 
\sum\limits_{\vec{n}} \delta_{y,x+\vec{n}}\delta_{t+1,t'}
+ \frac{\beta}{N_t}m_s \delta_{xy} \delta_{t+1,t'} & : t~\textrm{even}\\
\delta_{xy}\delta_{tt'}- \delta_{xy} \delta_{t+1,t'}
\exp(-i \frac{\beta}{N_t} \phi_{x,(t-1)/2})& : t~\textrm{odd}~ 
\end{array}
\right.\label{eq:fermionmatrix_2nd}
\ee
we can then rewrite the fermionic component of the action as
\be
S_F= S_\psi+S_\eta= \sum\limits_{t,t'=0}^{2N_t-1} \sum_{x,y}
\left(\psi^*_{x,t} M_{(x,t)(y,t')}\psi_{y,t'}+
\eta^*_{x,t} M^*_{(x,t)(y,t')}\eta_{y,t'} \right)~.
\ee
Note that the Hubbard-Coulomb field appears only on odd time slices here.
It hence follows that in HMC simulations the pseudofermion
fields have twice as many components as the Hubbard-Coulomb field.
Rederiving the fermion force (Eq.~(\ref{eq:fermforce_compact}))
for the 2nd order action is straightforward and yields
\begin{align}
(F_\chi)_{(x,t)}=-2 \frac{\beta}{N_t} \textrm{Im}\left[ \eta^*_{(x,2t+1)} 
e^{-i\frac{\beta}{N_t}\phi_{x,t} }\xi_{(x,2t+2)} \right]
\end{align}
Likewise, the order parameter for sublattice symmetry breaking
is computed as
\begin{align}
\langle \Delta_N \rangle
=\frac{-2}{N_tL_x L_y} \sum\limits_{t=0}^{N_t-1}
\Re\left\langle \sum\limits_{x \in x_A} M^{-1}_{(x,2t+1)(x,2t)}-
\sum\limits_{x \in x_B} M^{-1}_{(x,2t+1)(x,2t)}\right\rangle~.
\end{align}

\subsection{The partially screened Coulomb potential}
\label{subsec:partscreenedCoul}

We now turn to the interaction matrix $V$, which appears in the
Hamiltonian and thus in the Hubbard action
\be
S_\phi=\frac{\delta}{2}\sum_{t=0}^{N_t-1}\sum_{x,x'} \phi_{x,t}V_{xx'}^{-1} \phi_{x',t}~.
\label{eq:hubbact}
\ee
Note that we have absorbed 
the factor of $e^2$ ($=\alpha$ in our natural units)
in the definition of the matrix V here. Also note that it is the
\emph{bare} coupling constant $\alpha\approx 1/137$
that enters into $V$, not the
effective coupling constant $\alpha_\mathrm{eff} = \alpha/v_F \approx 300/137$
(with $v_f = 3\kappa a/2$, i.e.~the interaction strength relative to
the kinetic energy as described by the free hopping Hamiltonian $H_{tb}$).   
In the simulations we will account
for dielectric screening by a substrate through rescaling 
of the charge $e^2 \to e^2 / \epsilon$ and hence of the
potential. We discuss our results
both in terms of $\epsilon$ and in terms of the correspondingly 
screened \emph{effective} fine-structure constant 
$\alpha_{\mathrm{eff}}= \alpha_{\mathrm{eff},0}/\epsilon \approx 2.2 /\epsilon$.

It is clear
that all results will depend, perhaps strongly, on the
the two-body potential which is used and that
ultimately the validity of the entire study depends on 
making physically sound assumptions about the exact
form of $V$ (the fact that the inverse of $V$ appears will
be addressed below). It is obvious that $V_{xx'}$ should depend only
on the distance between the two lattice sites. Thus, we have
\be
V_{xx'}= V(x_1-x_1',x_2-x_2')\equiv V(r)~.
\ee
The straight-forward choice would be 
to assume that $V(r)$ is essentially a standard Coulomb potential 
$V(r)= e^2/r$, with a short-distance cut-off on the order
of half a lattice spacing (as discussed 
in Refs.~\cite{Brower:2012zd,Brower:2012ze}). 

It turns out, however, that this is insufficient to account for the
short-distance screening by the electrons in the $\sigma$-bands of
graphene and other higher energy states. To obtain a quantitatively
more accurate description of the partially screened Coulomb
interactions we use the results of Ref.~\cite{Wehling:2011df} where
this screening was investigated within a constrained random phase
approximation (cRPA). In particular, we use the  
numerical values from this reference for the effective strengths
of the on-site repulsion ($U_{00}=V_{xx}$), the nearest-neighbor
($U_{01}=V_{xx'}$ with $|x-x'|= a$), 
next-nearest-neighbor ($U_{02} = V_{xx'}$ with $|x-x'| = \sqrt 3 a $) and third-nearest-neighbor ($U_{03}= V_{xx'}$ with $|x-x'| = 2 a $)
interactions which are given by: 
\begin{center}
\vspace{1mm}
\begin{tabular}{|c|c|c|c|c|}
\hline
 & $U_{00}(\textrm{eV})$ & $U_{01}(\textrm{eV})$ & $U_{02}(\textrm{eV})$ & $U_{03}(\textrm{eV})$  \\
\hline
cRPA & $9.3$ & $5.5$ & $4.1$ & $3.6$\\ 
\hline
\end{tabular}
\vspace{1mm}
\end{center}
This part of the short-distance screening was implemented in
Ref.~\cite{Ulybyshev:2013swa} as well already.
For larger distances, however, it was then simply assumed that the
potential continues to fall off like $V(r)=e^2/(r\,\epsilon_0)  $,
with the constant $\epsilon_0$ fixed to $\epsilon_0=1.41$ so as to
match $V(2a)$ to $U_{03}$. 

As mentioned in the introduction already, there is no screening of the
long-range Coulomb tails in graphene, however. In order to describe
the screening from the $\sigma$-bands and other localized electron
states we therefore use a phenomenological model $\epsilon_\sigma(\vec
k)$  that describes a thin film of thickness $d$ with a dielectric
screening constant $\epsilon_1$. This model was adapted for graphene
by placing the two-dimensional sheet in the middle of the film in 
Ref.~\cite{Wehling:2011df} where it was found that an excellent fit to
the computed cRPA dielectric screening at longer wavelengths is
obtained for 
\begin{align}
\epsilon^{-1}_\sigma(\vec{k})=\frac{1}{\epsilon_1} 
\frac{\epsilon_1+1+(\epsilon_1-1) e^{-kd}}{\epsilon_1+1-(\epsilon_1-1) e^{-kd}}~,
\label{eq:dielectric1}
\end{align}
with $\epsilon_1=2.4$ and 
$d=2.8 \; \textrm{\AA}$ ($\approx 1.41\cdot 10^{-3}\, \mathrm{eV}^{-1}$).
This model smoothly connects the explicit short-distance screening with the
unscreened long-wavelenth limit as $\epsilon_\sigma(\vec k) \to
1 $ for $k\to 0$. 

\begin{figure}
\begin{center}
\resizebox{0.49\textwidth}{!}{%
\includegraphics{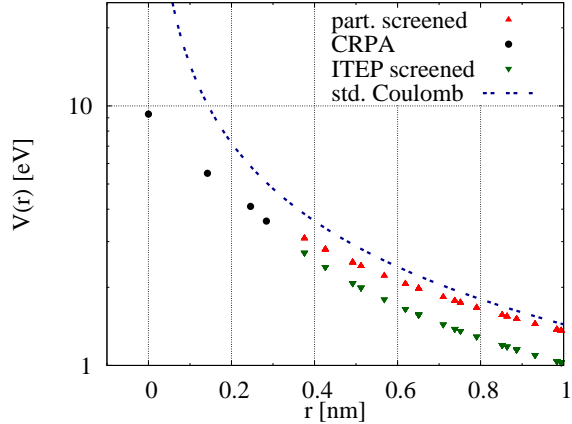}}
\end{center}
\caption{\label{fig:Vpotential1}
Comparison of the unscreened Coulomb potential (dashed blue),
the short-distance cRPA results of Ref.~\cite{Wehling:2011df} (black), the
long-range parts with constant $\epsilon_0$ to match $U_{03}$ as in
Ref.~\cite{Ulybyshev:2013swa} (green), and the long-range parts   
from  the phenomenological screening formula
of Ref.~\cite{Wehling:2011df} with unscreened Coulomb tails (red).}
\end{figure}

In order to obtain the corresponding partially-screened Coulomb
interaction matrix $V_{xx'} = V(r)$, first 
consider the unscreened potential $V_0(r)=e^2/r$ in two spatial
dimensions which in momentum space reads
$\widetilde{V}_0(\vec{k})=(2\pi e^2)/k$. The partially screended
$V(\vec r) $ is then obtained by the inverse Fourier transform with
the dielectric screening function $\epsilon_\sigma(\vec k)$ from
Eq.~(\ref{eq:dielectric1}) included,
\begin{align}
V(\vec{r})&=\frac{1}{(2\pi)^2}\int_{\mathbb{K}^2} d^2k\,
\widetilde{V}_0(\vec{k})\, \epsilon^{-1}_\sigma (\vec{k})
\, e^{-i\vec{k} \vec{r}}
= e^2 \int\limits_{0}^\infty dk\, \epsilon^{-1}_\sigma
(\vec{k})\,J_0(kr)\, ,
\label{eq:screenedV1}
\end{align}
where $J_0(x)$ is a Bessel function of the first kind. The
remaining one dimensional and well-behaved
integral can be easily computed numerically. 
Fig.~\ref{fig:Vpotential1} shows a comparison between the standard
Coulomb potential, the potential used in Ref.~\cite{Ulybyshev:2013swa}
(referred to as ``ITEP screened potential'' in the following) and the result
obtained from Eq.~(\ref{eq:screenedV1}) (referred
to as ``partially screened Coulomb potential''). 
It can be seen that the result from Eq.~(\ref{eq:screenedV1})
indeed connects with the explicit cRPA values at short distances and
smoothly approaches the unscreened Coulomb potential at large distances.

For portability, and to gain a better understanding how relevant screening
is on different length scales, we have parametrized $V(r)$. It turns
out that an Ansatz which assumes exponential (Debye) screening works well, 
provided that the mass (inverse screening length) is allowed to depend
on the length scale. 
Ultimately, for simulations we use a piece-wise defined potential
\begin{equation}
   V(r)  = \left\{ \begin{array}{lr}      
     U_{00},U_{01},U_{02},U_{03} & , \; r\le 2a \\
   e^2  \left( \big[ \exp(-m_2 r) /(m_1 r)^{\gamma}\big]m_0  +m_3
   \right)  &, \; r > 2a  
      \end{array}
          \right.\label{eq:potfit}
\end{equation}       
where $a$ is the lattice spacing, $m_1 = 1 $eV,  and the
parameters $m_0$, $\gamma$ , $m_2$, $m_3$ differ depending on the
length-scale as is summarized in the following table:
\begin{center}
\vspace{2mm}
\begin{tabular}{|c|c|c|c|c|}
\hline
 & $m_0$ [eV] & $\gamma$ & $m_2$ [eV] & $m_3$ [eV]  \\
\hline
$8a \ge r > 2a$ & $9.0380311$ & $0.632469$ & $144.354$ & $62.41496$  \\
\hline
$30a \ge r > 8a$ & $2.0561977$ & $0.862664$ & $27.8362$ & $15.29088$ \\
\hline
$120a \ge r > 30a$ & $1.03347891$ & $0.990975$ & $0.0$ & $-0.134502$  \\
\hline
$ r > 120a$ & $1.0$ & $1.0$ & $0.0$ & $0.0$  \\
\hline
\end{tabular}
\vspace{2mm}
\end{center}

We find that using these values our Ansatz differs from the numerical
results from Eq.~(\ref{eq:screenedV1}) by no more than $\sim 0.1\%$
for any $r$ and hence is sufficently accurate for our purposes.  
Furthermore, these parameters show that while there is rather strong screening
for $r<8a$, the Coulomb potential remains
essentially unscreened  for $r>30a$. 

Since in the simulation we want to maintain translational
invariance in order to minimize boundary effects, we have to make the
Coulomb interaction also translationally invariant. A naive sum over
infinitely many mirror charges with the long-range $1/r$ interaction
would not converge. In order to restore translational invariance in
the interaction while keeping the physical infrared cutoff that is
set by the size of a finite graphene sheet at the same time, we choose
to account for the boundary conditions in the following way: 
For any two points $x$ and $x'$ on the lattice, we determine
the shortest path between these points which can be constructed
if boundary crossings are allowed. The matrix element $V_{xx'}$
is then the potential $V(r)$ associated with this path. 
A potential constructed in such a way is translationally invariant,
but does not introduce infrared divergences. The natural infrared
regulator is then given by the finite size of the sample.

In Eq.~(\ref{eq:hubbact}) the inverse of $V$ appears. How to
efficiently compute this expression is a non-trivial technical problem.
Explicitly inverting $V$ and storing the elements of $V^{-1}$ requires
large amounts of memory. Instead, what is currently done in our
code is to use the conjugate gradient algorithm to invert $V$
on a given source $\phi$. This amounts to finding the solution $\eta $ of
\be
\phi_{x,t}=\sum_{x'} V_{xx'} \eta_{x',t}~ .
\label{eq:hubbact3}
\ee
To do this repeated computations of $V \eta $  are required. 
Since $V$ depends on the distance only, formally 
this is a two-dimensional convolution (for fixed $t$),
which suggests carrying out the computation in momentum space.
If the matrix $V$ is constructed as discussed above, always
using the shortest path to determine $V(r)$, Eq.~(\ref{eq:hubbact3})
can be brought into a form which is exactly a cyclic
convolution by introducing a rectangular fine-grid coordinate 
system as discussed in Sec. \ref{subsec:graphlat}
(the base vectors then coincide with the axes of
periodicity): We introduce
fields $\phi' $ and $\eta' $ on the fine grid.  The former has the
same value as $\phi $ on those points which are occupied by a carbon atom,
and is zero on all other points. 
Likewise, we introduce
a matrix $V'$ over the fine grid which contains $V_{x0}$
at each point. We then obtain 
$\widetilde{\phi}'_{k,t}$ (only the space-index is transformed)
and $\widetilde{V}'_k$ 
in this coordinate
system using the CUDA Fast Fourier Transform 
(cuFFT) library. $\phi $ in Eq.~(\ref{eq:hubbact3}) is then equal to the 
inverse Fourier transform of the point-wise product of
$\widetilde{V}'_k$ and $\widetilde{\eta}'_{k,t}$, restricted to points
which coincide with points on the hexagonal lattice. 

Note that we have first obtained the screened potential $V(r)$ by
numerically integrating Eq.~(\ref{eq:screenedV1}) 
\emph{in the continuum and infinite volume}. 
From this the discretized matrix $V$ is constructed. 
The discretized $\widetilde{V}'_k$ for
 the finite periodic lattice is then
obtained from this result. This procedure is free of 
additional sources of discretization and finite volume errors,
which would enter if one instead applied screening via $\epsilon^{-1}_\sigma(\vec{k})$
to a discretized $\widetilde{V}'_k$ which was obtained from
an unscreened matrix $V$.

Unfortunately it is not (at least not trivially) possible
to compute the inverse of $V$ directly in momentum space
by doing point-wise divisions of $\widetilde{\phi}'_{k,t}$ and
$\widetilde{V}'_k$. This is due to the additional spatial points of
the fine-grid which don't correspond to a site on the hexagonal lattice.

\section{Results}
\label{sec:results}
We simulate the interacting tight-binding theory via Hybrid-Monte-Carlo
for $L_x=L_y=18$ and $N_t=20$. We choose $\beta=2.0\,\mathrm{eV}^{-1}$ 
for the entire study.
The reasoning behind this choice of $\beta$ is the following: It was discussed
in Ref.~\cite{Armour:2011ff} that 
a temperature driven phase transition of the Berezinskii-Kosterlitz-Thouless 
type destroys sublattice symmetry breaking for high temperatures of the
electron gas. In Ref.~\cite{Buividovich:2012nx} the critical
temperature separating the high- and low-temperature phases
was estimated to be $T_c \approx 1.3 \cdot 10^4 \,\mathrm{K}$
(corresponding to $\beta \approx 0.89\, \mathrm{eV}^{-1}$).
Subsequently, both Refs.~\cite{Buividovich:2012nx} 
and \cite{Ulybyshev:2013swa} conducted simulations at 
$\beta=2.0\,\mathrm{eV}^{-1}$,
which corresponds to $T \approx 5.8 \cdot 10^3 \,\mathrm{K}$, well below $T_c$.
We adopt this choice to 
make a direct comparison possible. Ultimately, in order to
realistically account for finite temperature, lattice vibrations
must also be included, which is beyond the scope of our present work.
 
Our first goal is to reproduce the results of Ref.~\cite{Ulybyshev:2013swa}. 
We thus start by simulating the second order fermion matrix using
compact Hubbard field variables and construct
a screened potential accordingly (constant screening at long
distances, cRPA results at short distances). This is followed by
a simulation of the partially screened potential. The first order
discretization scheme is then used.

For both cases we choose different values
for the rescaling factor $\epsilon$ in the range
$\epsilon = 0.45 \ldots 1.0$ (which in fact mostly lie
outside of the physical region). 
This corresponds to values of the effective 
fine-structure constant $\alpha_\mathrm{eff}$ in the range 
$\alpha_\mathrm{eff} \approx 5.0 \ldots 2.0$. For each
choice of $\epsilon$ we simulate 
$m=0.5,0.4,0.3,0.2,0.1\, \mathrm{eV}$. 
We use a Sexton-Weingarten multiscale integrator with two scales:
The step-size for the Hubbard force is $\delta=1.0$ and for the
fermion force is $\delta/10=0.1$. The trajectory length
is $L=10$. We find that this yields satisfactory Metropolis
acceptance rates of $\sim 2/3$ for each choice of $m$ and
$\epsilon$. The number of conjugate gradient iterations required
for inverting the fermion matrix strongly increases with
decreasing mass. The smaller masses are thus substantially
more expensive than the larger ones in terms of computer time.

We begin with investigating thermalization times. Regardless
of the fact that the Hubbard-Coulomb field enters as a compact link variable
 in the fermion operator, it is a non-compact variable
in the Hubbard action. This means that unlike in theories
with strictly compact field variables (e.g. $SU(N)$ gauge theories), 
one can move arbitrarily
far away from thermal equilibrium. The choice of proper starting
conditions is therefore a non-trivial matter. It is complicated
by the fact that acceptance rates are strongly reduced
far from equilibrium. 

\begin{figure}
\begin{center}
\resizebox{0.49\textwidth}{!}{%
\includegraphics{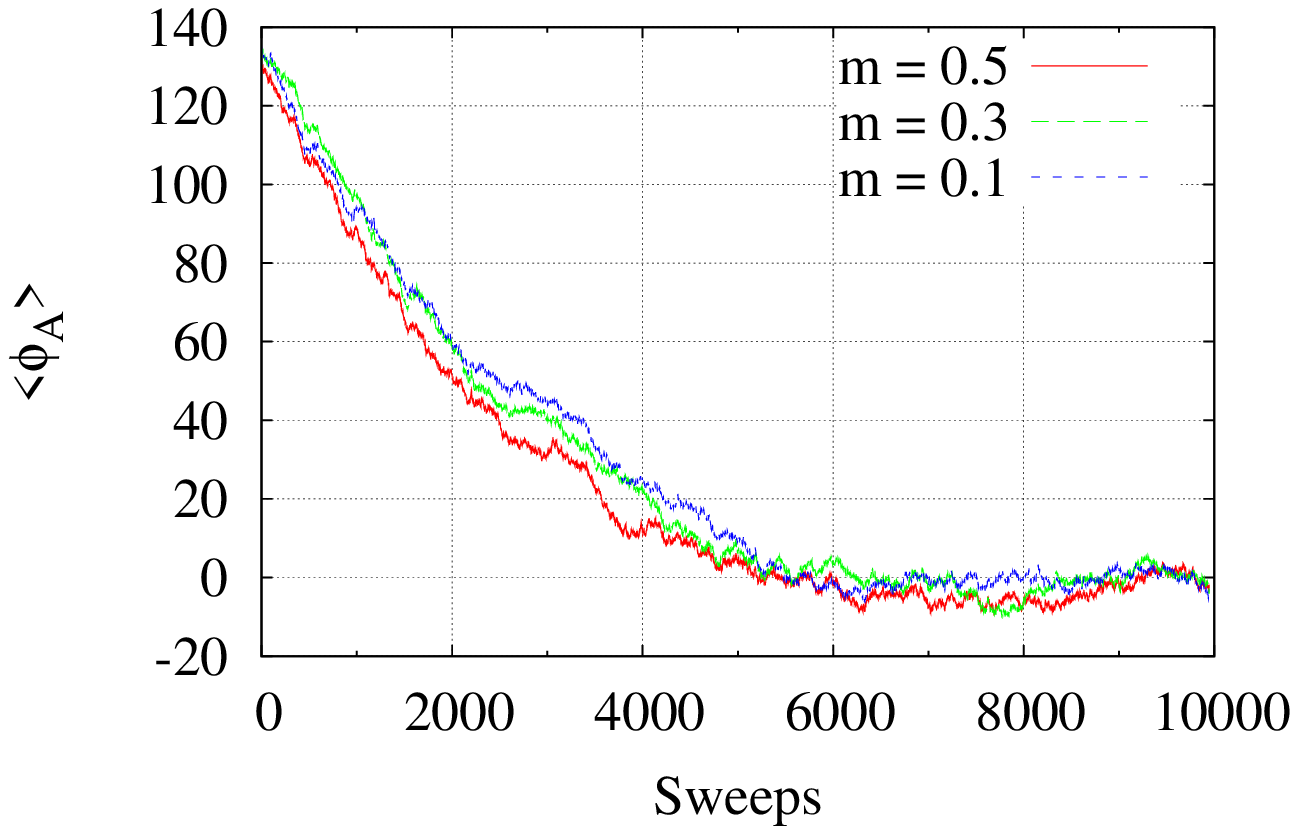}}
\resizebox{0.49\textwidth}{!}{%
\includegraphics{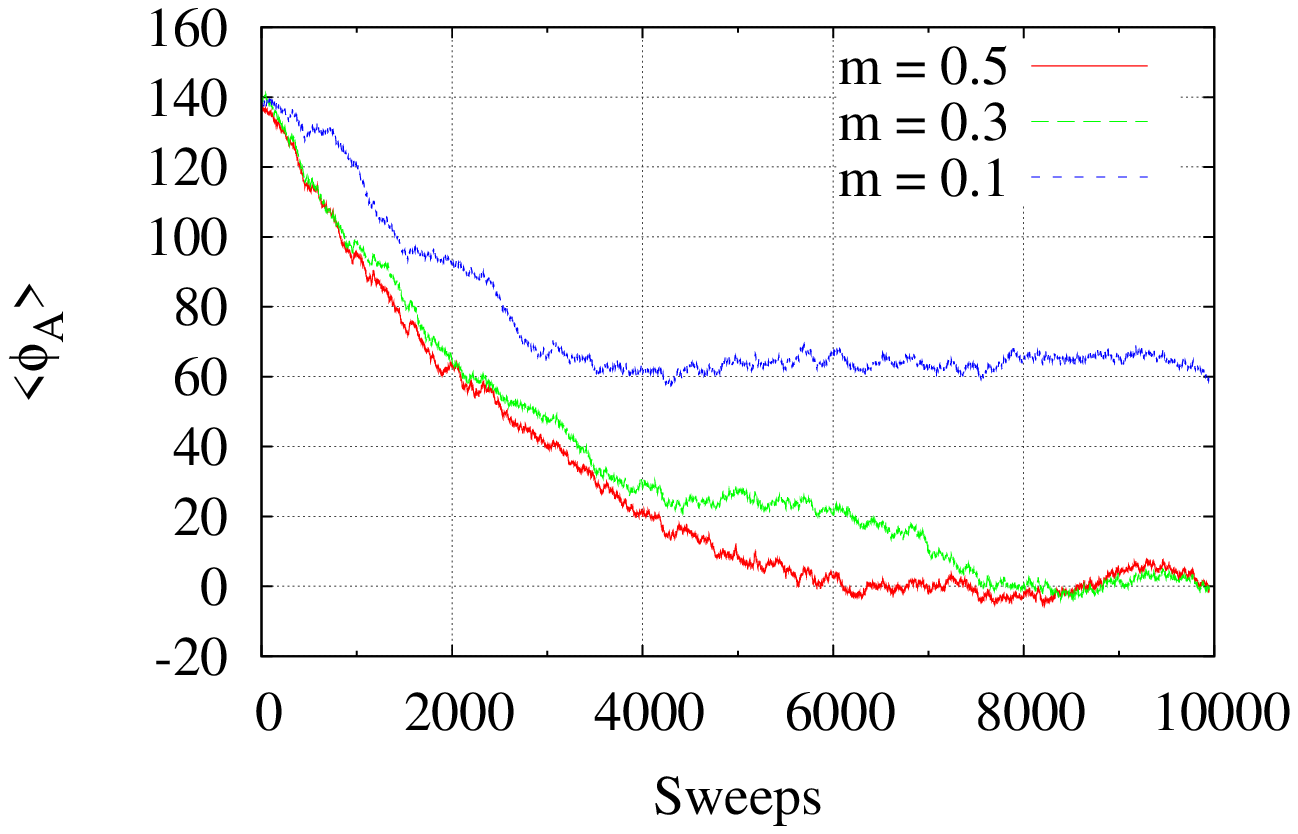}}
\end{center}
\caption{\label{fig:thermo_itep2ndorder}
Thermalization of the $\phi$ field on sublattice A for an ordered start 
(ITEP-screened potential). Left: $\epsilon=0.5$ ($\alpha_\mathrm{eff}\approx 4.379$). Right: $\epsilon=0.9$ ($\alpha_\mathrm{eff}\approx 2.433$)
. }
\end{figure}

For the first study (2nd order fermion matrix,
constant screening at long distances) we have done the following:
For each set of parameters we initialize the $\phi$ field
by setting $\phi_{x,t}=300$ for each $(x,t)$. We then
conduct $100$ MD trajectories without a Metropolis check. 
This does not suffice to bring the system into equilibrium,
but it brings the system into a region where Metropolis acceptance
is reasonable. After these trajectories we switch on the Metropolis check.
In Fig. \ref{fig:thermo_itep2ndorder} it is shown
how the $\phi$ field evolves on one sublattice
(in this case sublattice $A$) over the $10k$ subsequent trajectories
for $\epsilon = 0.5, 0.9$ ($\alpha_\textrm{eff}\approx 4.379,2.433$) and $m=0.1,0.3,0.5\,\mathrm{eV}$. 
Before taking measurements
we discard on the order of $10k$ trajectories for each combination of
parameters. As the figure illustrates, at small coupling and $m=0.1\,\mathrm{eV}$
this potentially introduces a systematic error due to incomplete 
thermalization (discarding two or three times as many trajectories
does not help. $\langle\phi\rangle \approx 0.0$ is not reached, even
after $\sim 30k$ trajectories). This issue
has only affected the two smallest
values of $\alpha_\mathrm{eff}$ which are simulated and leads to a slight 
over-estimation of the order parameter in the $m\to 0$ limit.
We further discuss this below. 

\begin{figure}
\begin{center}
\resizebox{0.49\textwidth}{!}{%
\includegraphics{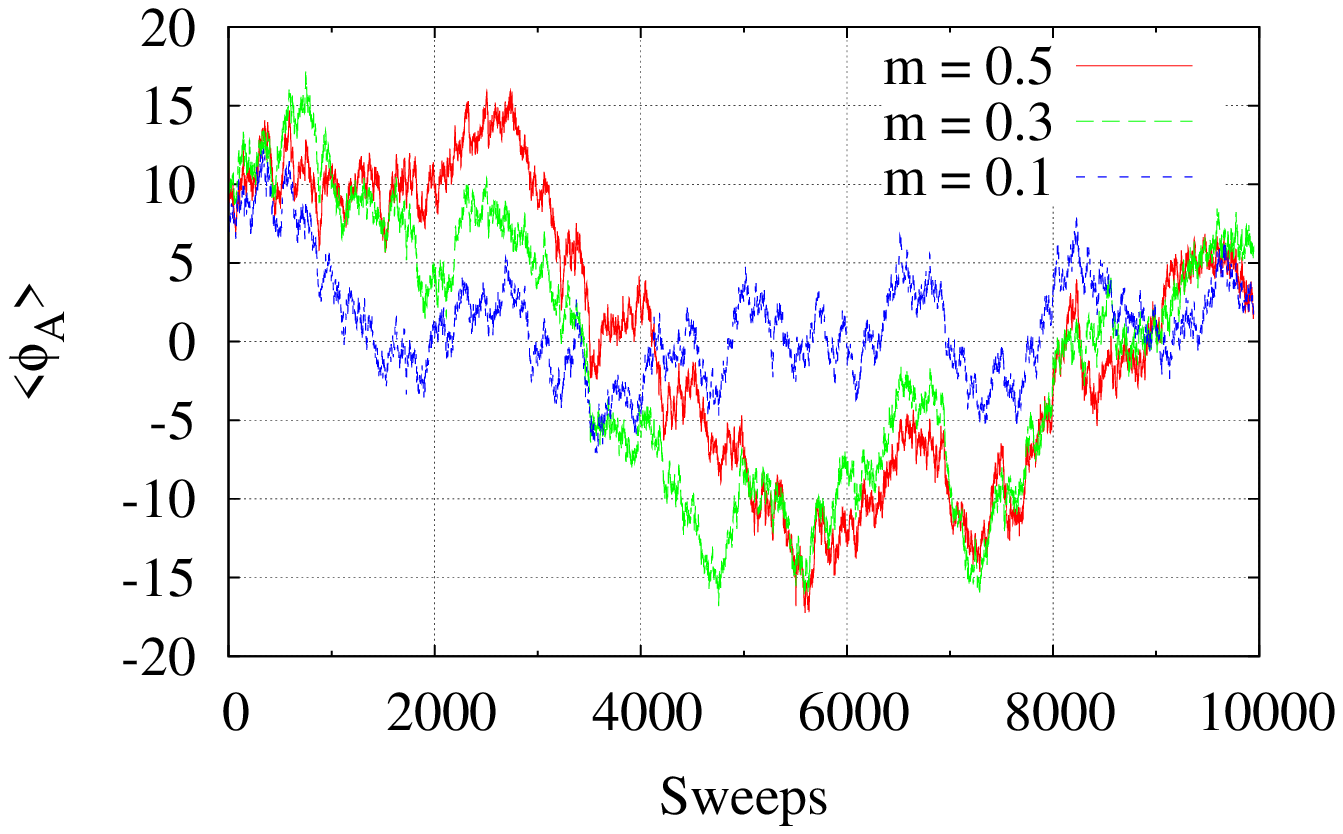}}
\resizebox{0.49\textwidth}{!}{%
\includegraphics{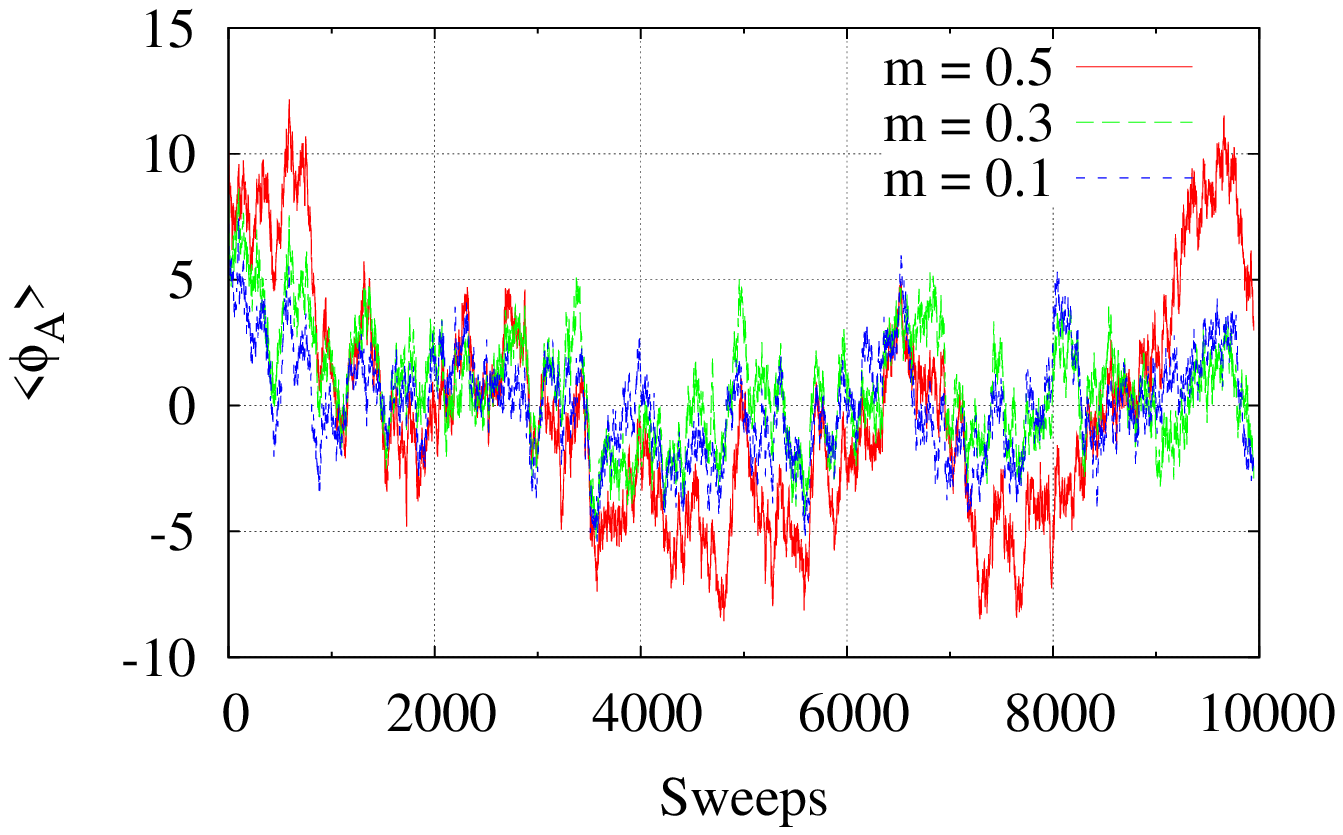}}
\end{center}
\caption{\label{fig:thermo_partscreen1storder}
Thermalization of the $\phi$ field on sublattice A when using a pre-thermalized
configuration of a different parameter set (partially screened potential). 
Left: $\epsilon=0.5$ ($\alpha_\mathrm{eff}\approx 4.379$). Right: $\epsilon=0.9$ ($\alpha_\mathrm{eff}\approx 2.433$)
.
}
\end{figure}

For the
simulations with the (correct) partially screened potential
we have revised our methodology and improved thermalization: We
first conduct one run of $10k$ trajectories using the largest mass
($m=0.5\,\mathrm{eV}$), which is cheap in terms of computer time. 
The final state of the $\phi$ field is then used
as starting condition for all other runs. We again start with
100 trajectories without Metropolis checks. 
In Fig. \ref{fig:thermo_partscreen1storder} the subsequent $10k$
trajectories are shown, again for the  $\alpha_\mathrm{eff}
\approx 4.379,2.433$ 
and $m=0.1,0.3,0.5\,\mathrm{eV}$. 
It appears that the system is already close to equilibrium after the $100$
initial iterations. We discard another $1000$ trajectories to be on
the safe side.

For both versions, we measure the order-parameter for sublattice symmetry
breaking $\langle \Delta_N \rangle$ using noisy estimators. $10$ source
vectors are used for each measurement. Measurements are done after every $10$th 
MD trajectory. We measure on several hundreds of independent configurations
(auto-correlations are estimated via binning) for each combination of $\alpha_\mathrm{eff}$
and $m$. We extrapolate the results to the $m\to 0$ limit by
doing a least squares fit to the form
\be
\langle \Delta_N \rangle(m)= c_1 m^2+c_2 m+c_3 \label{eq:mtozerofit}
\ee
\begin{figure}
\begin{center}
\resizebox{0.49\textwidth}{!}{%
\includegraphics{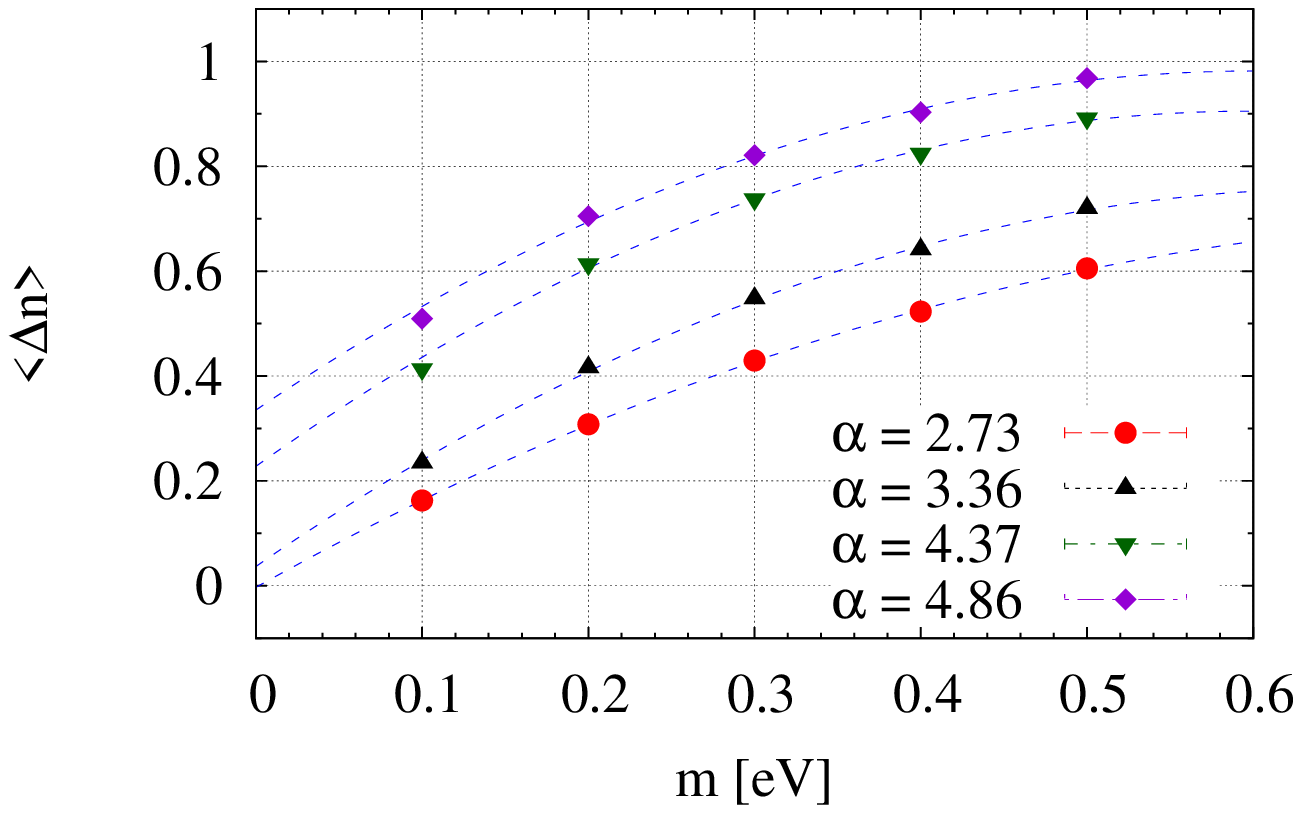}}
\resizebox{0.49\textwidth}{!}{%
\includegraphics{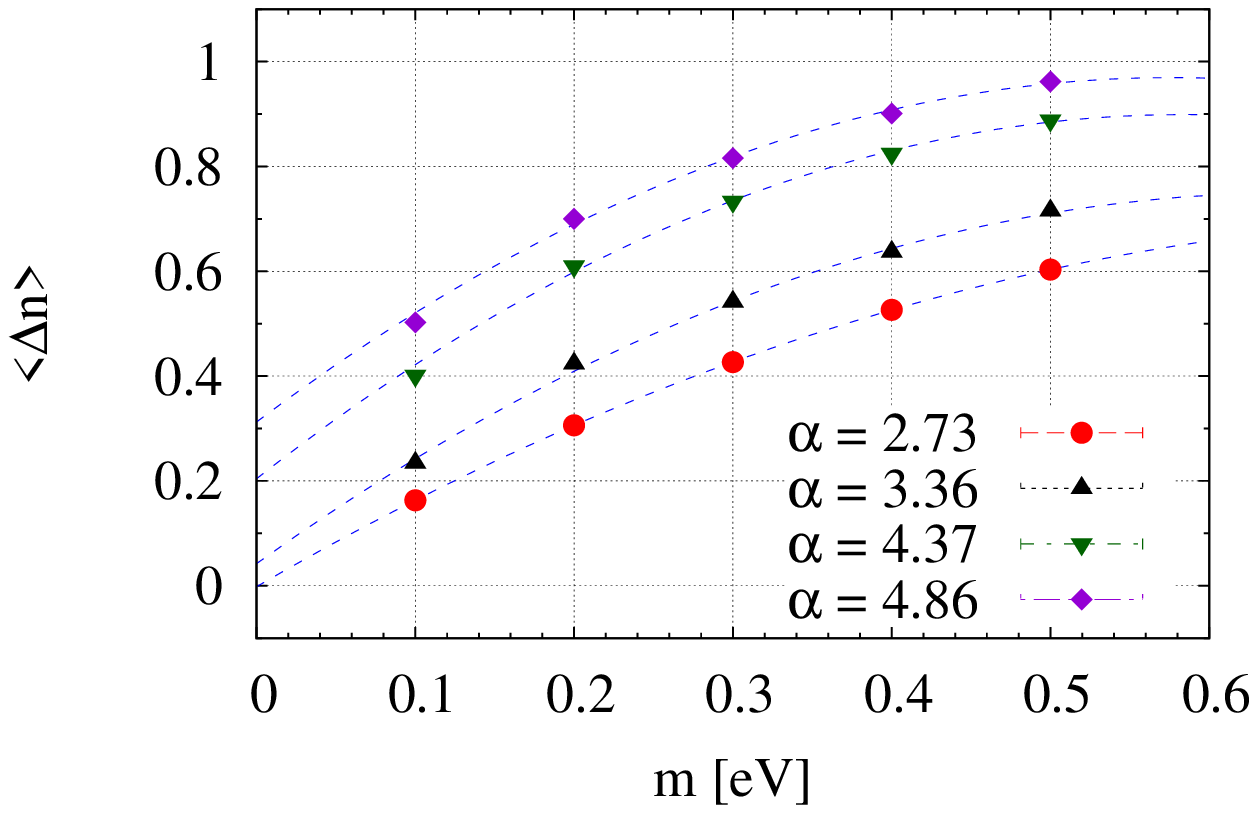}}
\end{center}
\caption{\label{fig:pbpmtozero}
Examples of $m\to 0$ extrapolation of order-parameter (error-bars are
smaller than points). 
Left: Partially screened
potential / 1st order discretization.
Right: Constant screening at long range / 2nd order discretization.}
\end{figure}
In Fig. \ref{fig:pbpmtozero} we show exemplary cases including
the fit via Eq.~(\ref{eq:mtozerofit}). The figure shows both versions of
the simulation (partially screened on the left and ITEP screened on the
right). This figure already suggests a strong similarity between the
two versions.
\begin{figure}
\begin{center}
\resizebox{0.49\textwidth}{!}{%
\includegraphics{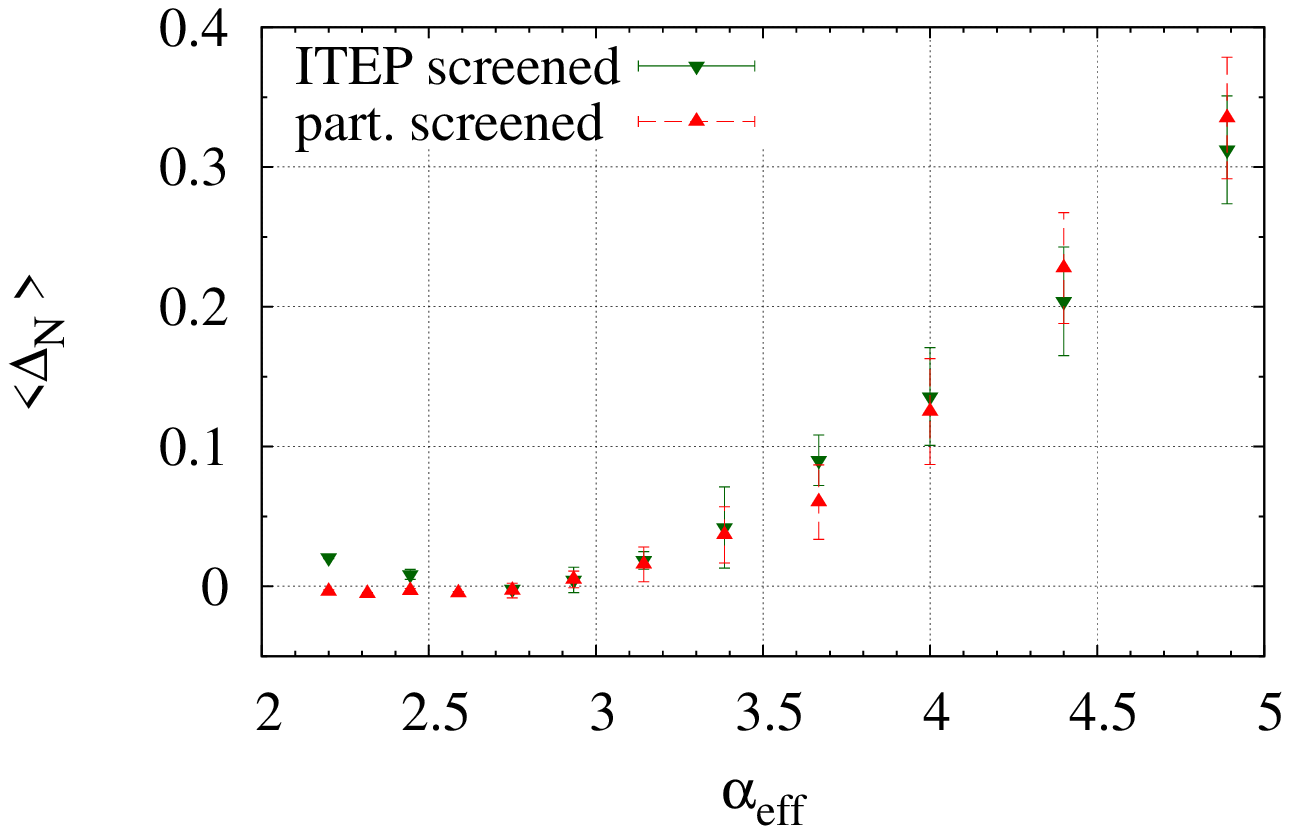}}
\resizebox{0.49\textwidth}{!}{%
\includegraphics{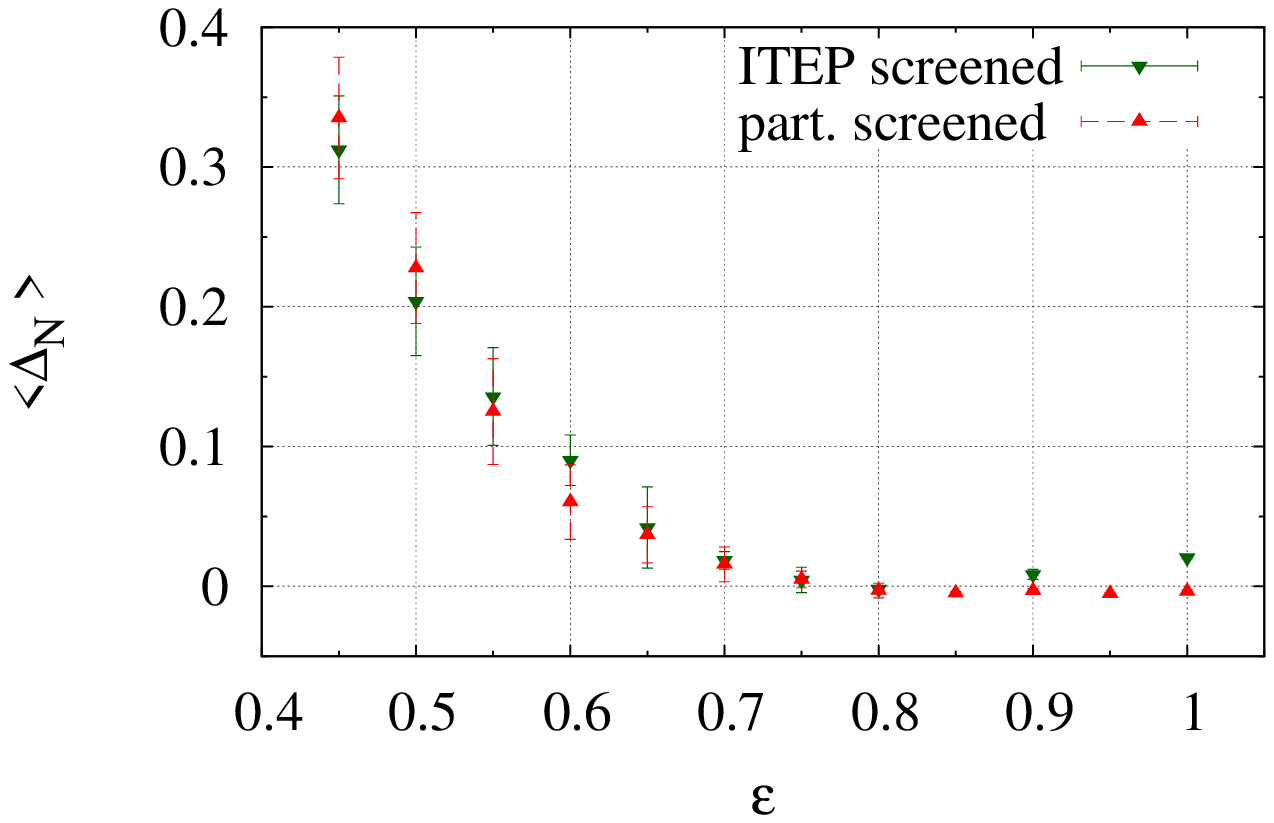}}
\end{center}
\caption{\label{fig:pbp1}
The $m\to 0$ extrapolated results for the order-parameter $\Delta_N$ 
as a function of $\alpha_\mathrm{eff}$ (left) and as a function of a rescaling parameter $\epsilon$ (right). }
\end{figure}
In Fig. \ref{fig:pbp1} we show the $m\to 0$ limiting cases for each $\alpha_\mathrm{eff}$. 
A strong similarity between the two different setups is observed.
In fact, both results within errors cannot be distinguished
from Ref.~\cite{Ulybyshev:2013swa}. Our findings are consistent
with a phase transition setting in around $\alpha_c\approx 3.14$. Note
that for the two smallest values of $\alpha_\mathrm{eff}$ we find 
$\langle \Delta_N \rangle$ slightly above zero for the first
setup, which is very likely
a systematic error due to the incomplete thermalization discussed
above.

\section{Conclusions and Outlook}
In this work we have conducted a Hybrid-Monte-Carlo
simulation of the interacting tight-binding model
for a $18 \times 18 \times 20$ lattice with periodic
boundary conditions. We have simulated
the system using two different setups: 
\begin{enumerate}[(a)]
\item{A second order fermion matrix and a potential which is
screened both at short and long distances.}
\item{A first order fermion matrix
and a partially screened Coulomb potential.}
\end{enumerate}
We have investigated spontaneous breaking of sublattice
symmetry in both cases. With our setup (a) we have
demonstrated consistency with Ref.~\cite{Ulybyshev:2013swa}.
We confirm the result of these authors, who find
that the phase transition occurs at $\alpha_c\approx 3.14$.
With our setup (b) we have investigated how this
result changes when the unscreened long-range Coulomb tails 
of the electronic interactions in graphene at half filling are
properly accounted for. We have demonstrated     
that the effect on the order parameter $\langle\Delta_N\rangle$
is small. We find, within errors, no difference between the
two setups, for the entire range of $\alpha_\mathrm{eff}$ which was
considered. This result suggests that development
of the band-gap is insensitive to the long-distance
part of the potential. It also confirms that screening of the
interactions by electrons in the $\pi$-orbitals may indeed be
a mechanism which explains the experimental finding
that suspended graphene is a conductor. 

For the future, it is immediately clear that these results
should be extrapolated to both
the infinite volume and the (time-like) continuum limit. In particular
it will be interesting to see whether the long-distance behavior
gets more important at larger volumes. Aside from this obvious
extension of the investigation presented in this paper, 
there is a large number of other open problems which are addressable
with our existing code, or slightly modified versions thereof.
We may study the effect of various open boundaries for instance.
It would also be possible to include additional terms in the Hamiltonian,
which describe phonon interactions or external magnetic fields
\cite{Gorbar:2002mf}. There are a number of possibilities which do not
introduce a fermion sign problem.  

Another problem of interest is to investigate
the effect of interactions on the topological neck-disrupting
Lifshitz transition which occurs in pure tight-binding models when the 
Fermi surface moves across a van Hove singularity in two dimensions,
where the density of states diverges, and which can be interpreted as
an excited state transition in the particle-hole excitation spectrum
as shown in Ref.~\cite{Dietz:2013sga}.
To study the role of interactions on this electronic transition,
 a chemical potential needs to be added. Unfortunately this creates a
fermion sign problem which cannot be trivially removed. For the immediate
future, however, we may add a spin-density chemical potential
without sign problem similar to the staggered spin-density mass term 
used in this work and analogous to finite isospin chemical potential
in QCD. One could then study the divergence in the corresponding
susceptibility with finite-size scaling and see
in what way the logarithmic divergence indicative of the
neck-disrupting Lifshitz transition in two dimensions is modified by
the interactions. Ultimately it would also be interesting to
investigate whether the tight-binding model with interactions 
for the electronic excitations in graphene happens to belong
to the class of theories in which a fermion sign problem can be dealt
with in one way or another.

\section*{Acknowledgments}

We have benefited from discussions with Richard Brower, Pavel Buividovich, Maxim
Ulybyshev and the late Mikhail Polikarpov. We thank 
Manon Bischoff and Michael K\"orner for proof-reading
the manuscript. Also, we thank Peter Neuroth for
creating the illustration of the graphene lattice. 

This work was supported by
the Deutsche Forschungsgemeinschaft within 
SFB 634, by the Helmholtz International Center for FAIR within the
LOEWE initiative of the State of Hesse, and the European Commission,
FP-7-PEOPLE-2009-RG, No. 249203.  All results were obtained using
Nvidia GTX and Tesla graphics cards on the Scout Cluster of the
Center for Scientific Computing (CSC) of the University of Frankfurt
and on the Lichtenberg Cluster of the Hochschulrechenzentrum
Technische Universit\"at Darmstadt.

\end{document}